\documentclass[journal]{IEEEtran}
\usepackage{graphicx}
\usepackage{url}
\usepackage{amsmath, amsfonts}
\usepackage{amsthm, amssymb}
\usepackage{comment}
\usepackage{xcolor}
\usepackage[font=small]{caption}
\usepackage{subcaption}
\usepackage{array}
\usepackage{comment}
\usepackage{cite}

\usepackage{pifont}
\newcommand{\xmark}{\ding{55}}

\DeclareMathAlphabet\mathbfcal{OMS}{cmsy}{b}{n}

\ifCLASSINFOpdf
\else
\fi
\begin{document}

\title{Multi-Modality Sensing in mmWave Beamforming for Connected Vehicles Using Deep Learning}

\author{Muhammad Baqer Mollah,~\IEEEmembership{Graduate Student Member,~IEEE,}
        Honggang Wang,~\IEEEmembership{Fellow,~IEEE,}
        \\ Mohammad Ataul Karim,~\IEEEmembership{Life Fellow,~IEEE,}
        and Hua Fang,~\IEEEmembership{Senior Member,~IEEE}%

\thanks{Muhammad Baqer Mollah and Mohammad Ataul Karim are with the Department of Electrical and Computer Engineering, University of Massachusetts Dartmouth, MA 02747 USA (Emails: mmollah@umassd.edu; mkarim@umassd.edu). Honggang Wang is with the Department of Graduate Computer Science and Engineering, Katz School of Science and Health, Yeshiva University, NY 10016, USA (Email: honggang.wang@yu.edu). Hua Fang is with the Department of Computer and Information Science, University of Massachusetts Dartmouth, MA 02747, USA (Email: hfang2@umassd.edu). (Corresponding Author: Honggang Wang)}
\thanks{The preliminary version of this work has been presented at IEEE International Conference on Communications 2024 \cite{mollah2024position}.}
\thanks{This research is partly supported by National Science Foundation (NSF) under the grants 2010366 and 2140729.}
\thanks{This article has been accepted for publication in IEEE Transactions on Cognitive Communications and Networking. This is the author's version which has not been fully edited and content may change prior to final publication.}
\thanks{DOI: https://doi.org/10.1109/TCCN.2025.3558026}
\thanks{\copyright 2025 IEEE. Personal use is permitted, but republication/redistribution requires IEEE permission. See https://www.ieee.org/publications/rights/index.html for more information.}%
}

\markboth{IEEE Transactions on Cognitive Communications and Networking}%
{Shell \MakeLowercase{\textit{et al.}}: Bare Demo of IEEEtran.cls for IEEE Journals}

\maketitle

\begin{abstract}
    Beamforming techniques are considered as essential parts to compensate severe path losses in millimeter-wave (mmWave) communications. In particular, these techniques adopt large antenna arrays and formulate narrow beams to obtain satisfactory received powers. However, performing accurate beam alignment over narrow beams for efficient link configuration by traditional standard defined beam selection approaches, which mainly rely on channel state information and beam sweeping through exhaustive searching, imposes computational and communications overheads. And, such resulting overheads limit their potential use in vehicle-to-infrastructure (V2I) and vehicle-to-vehicle (V2V) communications involving highly dynamic scenarios. In comparison, utilizing out-of-band contextual information, such as sensing data obtained from sensor devices, provides a better alternative to reduce overheads. This paper presents a deep learning-based solution for utilizing the multi-modality sensing data for predicting the optimal beams having sufficient mmWave received powers so that the best V2I and V2V line-of-sight links can be ensured proactively. The proposed solution has been tested on real-world measured mmWave sensing and communication data, and the results show that it can achieve up to $98.19$\% accuracies while predicting top-$13$ beams. Correspondingly, when compared to existing been sweeping approach, the beam sweeping searching space and time overheads are greatly shortened roughly by $79.67$\% and $91.89$\%, respectively which confirm a promising solution for beamforming in mmWave enabled communications.
\end{abstract}

\begin{IEEEkeywords}
Beamforming, Connected and Autonomous Vehicles, Deep Learning, Millimeter-Wave Communications, Multi-Modality Sensing, Out-of-Band Information.
\end{IEEEkeywords}
    
\IEEEpeerreviewmaketitle
    
\section{Introduction}    
    \IEEEPARstart{T}{he} utilization of millimeter-Wave (mmWave) bands (e.g., 28 GHz and 57-71 GHz bands) has been shown to bring an abundance of spectrum resources, considered as a key enabler for communications technologies in connected and autonomous vehicles (CAVs) domain \cite{mollah2024mmwave, xie2024learning, feng2024online}. For example, in the context of cooperative perceptions such as in high-throughput and low latency demanding applications in CAVs, the connected vehicles exchange a vast amount of 3D point cloud data from LiDAR sensors instead of lightweight textual data \cite{lin2024v2vformer, ngo2023cooperative}. Realizing the high throughput and low latency benefits, mmWave communications enabled by vehicle-to-infrastructure (V2I) and vehicle-to-vehicle (V2V) communications is expected to pave the way toward accessible and safe autonomous transportation systems.
    
    On the other hand, an inherent property of mmWave band is having high path attenuation of its signals, whereby massive antenna arrays are typically employed, drastically impacts quality performance degradation \cite{osa2023measurement,sun2018propagation}. Beamforming techniques may be utilized to address this bottleneck, including formulating the narrow beams, thereby achieving the desired high throughput from mmWave communications. Likewise, the narrow beams are expected to be aligned precisely and required to re-direct in accordance with the environmental settings and any changes. Generally, in codebook-based beamforming \cite{xue2024ai}, beam selection techniques are applied to find the optimal beams in a number of pre-defined beam codebooks.
    
    In practice, with standard defined beam selection techniques, the vehicles usually select the best beam pairs through an exhaustive beam measurements process \cite{xue2024survey}, which in turn introduces computing and latency overheads in applications involving highly dynamic moving vehicles. This overhead problem occurs mainly due to tight contact times (the time period the receiver received the correct packets), frequent beam realignments, and changes in channel state estimation to perform beam computing. Hence, it is crucial to avoid beam misalignment along with reduced and reasonable overheads in order to take the fully potential benefits from mmWave communications. For instance, deployment based the 3GPP 5G-NR (New Radio) and IEEE 802.11ad standards \cite{salehi2022deep} might take around $25$ and $19$ milliseconds, respectively to select the best beams through searching over all beam directions for $60$ beam pairs and need to run over repeatedly when the vehicles move forward.
    
    Considering this challenge in vehicular settings, recent works have suggested effective approaches to configure the communications links by leveraging the out-of-band contextual information/side information. As such, the out-of-band contextual information has been applied in mmWave beamforming tasks by utilizing other lower bands (sub-6 GHz bands) \cite{zhao2024lstm, huang2024sub, he2024efficient, xia2024millimeter, yashvanth2024impact, vuckovic2023paramount}, RADAR communications bands \cite{zhou2024radar, al2024enhancing, graff2023deep, luo2023millimeter}, or extracted useful sensing information from on-board vehicular non-RF sensing devices (i.e., position coordinates \cite{morais2024localization, xing2023location, rezaie2022deep}, visual data \cite{kim2024vomtc, hua2023computer, xu2023computer, huang2024image}, and point clouds \cite{nerini2023overhead, jiang2022lidar, mashhadi2021federated, klautau2019lidar}) are the few notable example works. According to \cite{roy2023going}, several approaches pursuing efficient beamforming have been introduced based on multi-modality sensing over the recent years. However, among these related research, most of the works are limited primarily to vehicle-to-infrastructure (V2I) communications, based on ray-tracing simulations, or considered single sensing modality, but utilizing real-world measured wireless data as well as considering the mobility of connected vehicles under V2V connectivity have not been investigated yet. Accordingly, we introduce a unified solution for mmWave beamforming for both V2I and V2V communications, unlike prior works, with the aid of a proposed multi-modal deep learning model. The main contributions of this work are listed in detail below:
    
    \begin{itemize}
    \item Drawing inspiration from the recent success of multi-modality sensing, we propose a 5G-NR standard compatible deep learning-based solution to predict a subset of beams, that is, top-$M$ beams, thereby significantly reducing the searching space and latency of beam sweeping.
    
    \item Instead of single modality, we employ different sensing modalities with a common goal on providing more detailed information and robustness in terms of errors and missing information beyond the limitations of the individual sensing in practical situations. Specifically, the position information is leveraged to facilitate on identifying the targeted vehicles in urban scenarios.
    
    \item Unlike other approaches, the proposed solution fully leverages the sensing data obtained from sensors, commonly available on the modern vehicles and road-side units, subsequently lowering down the complicacy.
    
    \item The proposed solution is validated with experiments on real-world multimodal 60 GHz mmWave wireless data, and present the results in terms of two meaningful matrices, namely accuracies and received power ratios to demonstrate its efficacy and applicability in beamforming tasks in reality.
    
    \item The proposed solution is shown to integrate into existing 5G-NR standard and the benefits we can get after integration. In particular, the results show that the proposed solution can reduce the beam sweeping searching and time overheads by $79.67$\% and $91.89$\%, respectively for top-13 beams.
    \end{itemize}
    
    \textit{Paper Organization:} The rest of this paper is organized as follows. In section II, we review the most relevant related works on efficient beamforming, in particular, that have been introduced in the literature. In section III, we present the considered system model and formulated problem. The detailed steps of the proposed multi-modal solution is introduced in section IV. Next, the multi-modal feature extraction architecture including its necessary components are described in section V. The experimental settings and results are presented in section VI. Finally, conclusions are summarized in section VII. The list of major notations and their descriptions used in this paper is summarized in Table \ref{tab: notations}.

\begin{table}[ht]
    \centering
    \caption{Summary of major notations and descriptions.}
    \label{tab: notations}
    \begin{tabular}{m{1.6cm}|m{5.5cm}}
    
    \hline \hline

    \textbf{Notation}   &   \textbf{Description} \\
    \hline

    $v_1$, $v_2$ \& $v_3$    &    Moving vehicles \\
    \hline

    $\mathcal{A}$    &    gNodeB, base station, or road-side unit \\
    \hline
     
    $N_R$     &    The numbers of antenna array elements \\
    \hline

    $\mathbf{h}_n $     &    Channel vectors in each communication links \\
    \hline
    
    $\mathbfcal{Q}$     &    Beam codebooks \\
    \hline

    $\mathcal{I}$    &    Unique beam indices \\
    \hline

    $\mathcal{B}$    &    The set of recommended beams \\
    \hline

    $X_{pos}$    &    Position data matrix \\
    \hline

    $X_{vis}$    &    Visual data matrix \\
    \hline

    $X_{lid}$    &    3D point cloud data matrix \\
    \hline

    $\mathcal{D}_s$    &    Training dataset \\
    \hline

    $\psi(\cdot, w)$    &    Neural network \\
    \hline

    $w$    &    Trainable weights of $\psi(\cdot, w)$ \\
    \hline

    $\psi_{w^*}$    &    Parameterized by optimal weights $w^*$ \\
    \hline


    $\hat{\mathcal{P}_t}$     &    Received power from predicted beams \\
    \hline
    
    $\mathcal{P}_t^{gt}$      &    Received power from ground-truth beams \\
    \hline

    $\mathcal{T}_{nr}$     &    Required time of beam selection in 5G-NR  \\
    \hline
    
    $T_{ssp}$     &    Single burst in synchronizing signal blocks \\
    \hline

    $t_{burst}$     &    Synchronizing signal burst periodicity \\
    \hline

    $\mathcal{T}_{nr}^{cnn}$     &    Sweeping time for the predicted beams \\
    \hline
    \end{tabular}
\end{table}

\section{Related Works}
    In this section, we review the most relevant works on efficient beamforming based on the out-of-band information which have been introduced in the literatures as follows.
    
\subsection{Uni-Modal Sensing Data}
    Recent works have shown that positioning, visual, and 3D point clouds sensing information have emerged as potential enablers for mmWave beamforming. The solution in \cite{morais2023position} has leveraged machine learning technique to investigate how position information from GPS sensors can help to reduce the beam training overheads. However, considering the localization errors of GPS data, the authors in \cite{ha2024radio} have proposed a beam alignment approach, where Bayesian approach has been applied on the radio maps at a geographical area's each spatial location. Meanwhile, the work in \cite{meng2024tdoa} has developed a data driven method by transformer networks with particular attention mechanisms to estimate the line of sight (LoS) time of arrival (ToA), and then, a model-driven method to locate the users in mmWave multiple-input multiple-output (MIMO) communications by approximating the maximum likelihood time difference of arrival (TDoA) with the estimated results. Likewise, the work in \cite{rezaie2022deep} emphasizes on both location and orientation information to perform 3D beam selection by utilizing deep learning technique.
    
    Similar concepts have been also utilized in \cite{mattos2022geolocation} and \cite{khosravi2022location}. In \cite{mattos2022geolocation}, the authors utilize spatially indexed historical information so that statistically significant beam sectors can be chosen in specific locations. Their proposed work is particularly focused on IEEE 802.11ad standard with an aim of reducing the time of beam sweeping, thereby improving the available time for data communications. And, the work in \cite{khosravi2022location} uses the information about spatial correlation of stronger channels in-between the receiver and sender at given locations so that the users are able to track the information beforehand to establish communication link quickly. Earlier, the authors in \cite{va2017inverse} had proposed an approach, where a multi-path fingerprint-based database is maintained which records fingerprints of necessary location information along with channel characteristics so that the knowledge of reliable beamforming can be known in advance.
    
    In addition to the utilization of position information, the capabilities of scene understanding by visual data by RGB or depth (RGB-D) sensing cameras have also been utilized in a number of studies. For instance, the authors in \cite{ahn2024sensing} and \cite{al2023intelligent} have presented computer vision assisted mobility management techniques. In particular, Anh et al. \cite{ahn2024sensing} have been utilized visual sensing information to understand the moving user and potential blockages by extracting the distance and azimuth/elevation angle like geometric channel parameters, which later have been used to predict the downlink state (i.e., data rate) using a deep neural network. The simulation results show that their proposed work outperforms the beam sweeping based on the 5G-NR standard-defined technique by achieving $30$\% more gains in throughput while ensuring a seamless beamforming gain over the targeted area. Whereas, the authors in \cite{al2023intelligent} have used computer vision techniques to understand the location and speed of the vehicles, at the same time, the potential obstacles, and later, this information has been utilized to determine the best handover decision. Here, the authors have incorporated a multivariate regression technique to play a vital role to predict the time until the moving vehicle arrives at the blockage area.
    
    Meanwhile, the authors in \cite{xiang2023computer} have designed a different computer vision assisted mmWave beamforming focusing on beam weight prediction, where the visual detection by a convolutional neural network and limited wireless feedback have been fused to enhance the spectral efficiency as well as a codeword rotation method implemented by Householder transformation to reduce the feedback overheads. Indeed, Osman et al. \cite{osman2023vehicle} have shown how visual sensing data collected from $360^{\circ}$ cameras can help to contribute on beam predictions in V2V communication scenarios in their work. Besides, in \cite{charan2023user}, the authors have focused on user identification task together with a deep neural network and bounding box estimation so that the computer vision aided communications can be employed in realistic crowded scenarios. 
     
    The work in \cite{lin2024multi} has presented a multi-task learning module to predict the optimal road side units (RSU) and beam pairs by proposing deep learning based proactive RSU selection network (PRSN) and beam pair searching network (BPSN), respectively for dynamic V2I settings. Here, the authors have utilized multiple cameras installed on the RSUs for the purpose of getting multiple views, which essentially enriches the visual sensing information in terms of making more accurate and reliable decisions than single view techniques. However, utilizing raw visual sensing data directly introduces considerable storage and computational overheads. Instead, the authors in \cite{imran2023environment,imran2024environment} and \cite{wen2023vision} have utilized targeted object masks and bounding boxes like environment semantics information to address such overheads problems.
    
    Although the aforementioned works have shown that the visual sensing data has been utilized as out-of-band information along with satisfactory scene understanding, having dark or glare visual images may lead to inaccurate beamforming decisions by the models/algorithms. In contrast, 3D structures captured by LiDAR point clouds can offer a better understanding of surrounding environment and scene understandings through high-resolution mapping and positioning. Accordingly, the authors in \cite{klautau2019lidar} and \cite{mashhadi2021federated} have considered utilizing point clouds for detection of line-of-sight, and then, predict the best beams by deep convolutional neural network and federated learning, respectively. While these two works have focused on predicting the current beams, similar work has been studied in \cite{jiang2023lidar} using by a recurrent neural network (RNN), and shown that with the aid of point clouds, it is possible to predict the future beams along with current ones.

    Wu et al. \cite{wu2023proactively} have shown in their work that leveraging convolutional neural network on point clouds and in-band mmWave signals can be a help in predicting the occurrence of line-of-sight link blockages in advance so that beam switching/hand-off decisions can be made to ensure reliable communications. Most importantly, the authors have validated their work on real-world datasets, showing that their approach can achieve around $95$\% and $80$\% accuracies while predicting the future blockages within $100$~ms and $1,000$ ms, respectively. Besides, Ohta et al. in \cite{ohta2023point} have demonstrated with real experiments that it is possible to leverage the inherent properties of point clouds to understand the wireless propagation environment. They have designed a proposed 3D convolutional neural network with long short-term memory and gradient boosting decision tree for implementing on point clouds so that the link quality, more specifically, the mmWave received signal attenuations due to any blockages (e.g., pedestrians) can be predicted proactively. Further, the work in \cite{pradhan2024} has introduced a deep learning based technique, namely COPILOT, to identify the best road side unit having high bandwidth mmWave connectivity for cooperative perception purposes. The aim of this COPILOT is to support connected vehicles to recognise the transient blockages along the road with the help of point clouds, which eventually can help them to decide connecting the best road side unit by proactive handoffs. In particular, when compared to traditional reactive handoffs techniques, the experiment with real $60$ GHz devices has shown that the COPILOT can offer upto $69.8$\% improvement on latency.

\begin{table*}[ht]
    \centering
    \caption{Summary of relevant state-of-the-art multi-modal sensing related works and our contributions.}
    \label{tab: comparison}
    \begin{tabular}{m{.6cm}|m{1.16cm}|m{.9cm}|m{.9cm}|m{.9cm}|m{1.75cm}|m{1.7cm}|m{.55cm}|m{.55cm}|m{1.6cm}|m{1.7cm}}
    
    \hline \hline
    \textbf{Work} & \textbf{Frequency Band} & \textbf{Position} & \textbf{Visual} & \textbf{Point Clouds} & \textbf{Problem} & \textbf{Highlighting Method} & \textbf{V2I} & \textbf{V2V} & \textbf{Validation} & \textbf{Dataset and tool} \\
    \hline \hline
    
    \cite{charan2022vision} & 60 GHz & \checkmark & \checkmark & \xmark & Beam selection & Computer vision & \checkmark & \xmark & Real-world data & DeepSense6G \\
    \hline

    \cite{raha2024advancing} & 60 GHz & \checkmark & \checkmark & \xmark & Beam selection & Transformers network & \checkmark & \xmark & Real-world data & DeepSense6G \\
    \hline

    \cite{zecchin2022lidar} & 28 GHz & \checkmark & \xmark & \checkmark & Beam selection & Deep neural network & \checkmark & \xmark & Synthetic raytracing data & Raymobtime \\
    \hline

    \cite{bian20243} & 60 GHz & \checkmark & \xmark & \checkmark & Beam selection & Deep neural network & \checkmark & \xmark & Real-world data & DeepSense6G \\
    \hline

    \cite{niu2023multi} & 28 GHz & \checkmark & \checkmark & \checkmark & Beam selection & Deep neural network & \checkmark & \xmark & Synthetic raytracing data & Raymobtime \\
    \hline

    \cite{patel2024harnessing} & 28 GHz & \checkmark & \checkmark & \checkmark & Multi-user beam selection & Deep neural network & \checkmark & \xmark & Synthetic raytracing data & Raymobtime \\
    \hline

    \cite{salehi2024flash} & 60 GHz & \checkmark & \checkmark & \checkmark & Sector selection & Federated learning & \checkmark & \xmark & Real-world data & FLASH \\
    \hline

    \cite{salehi2024multiverse} & 60 GHz & \checkmark & \checkmark & \checkmark & Beam selection & Digital twin & \checkmark & \xmark & Real-world data & FLASH and RemCom \cite{remcom} \\
    \hline

    \textbf{Ours} & \textbf{60 GHz} & \checkmark & \checkmark & \checkmark & \textbf{Beam selection} & \textbf{Deep neural network} & \checkmark & \checkmark & Real-world data & \textbf{DeepSense6G} \\
    \hline   

\end{tabular}
\end{table*}
    
\subsection{Multi-Modal Sensing Data}
    Beyond the single modality works, several solutions on leveraging muti-modality sensing to facilitate mmWave beamforming related tasks have been explored in recent years. For example, the works in \cite{charan2022vision} and \cite{raha2024advancing} have focused on deep learning assisted fusion of positioning and visual sensing data. In particular, the work \cite{charan2022vision} utilized fine-tuned ResNet-50 model and data normalization technique on visual data and position data, respectively before feeding into a concatenation module, and then used a multi-layer perceptron network to get the predicted beam indices. However, the work \cite{raha2024advancing} has incorporated transformer network and semantic localization techniques to contribute more on developing a robust beamforming approach while obtaining the visual data in varrying weather such as rainy, foggy, sunny, and snowy conditions. On the other hand, Zecchin et al. \cite{zecchin2022lidar} have proposed a fusion between point clouds and GPS position data, where the authors have improved the beam prediction accuracies by utilizing non-local attention block particularly for non-line-of-sight scenarios, and also utilized a loss function motivated by knowledge distillation as well as a curriculum training strategy to further improve the training speed as well as final predicted results. Finally, this work has been validated with a synthetic ray-tracing dataset, namely Raymobtime\footnote{\url{https://www.lasse.ufpa.br/raymobtime}}. In another work, Bian et al. \cite{bian20243} have presented a beam tracking method on fusing the measured point clouds and position data by a proposed deep neural network, however, the studies was limited to only beam tracking in vehicle-to-vehicle connectivity.

    Meanwhile, multiple works in \cite{niu2023multi, patel2024harnessing, salehi2024flash, salehi2024multiverse} have demonstrated how to fuse features from visual, point clouds, and position sensing data together. In particular, the work in \cite{niu2023multi} has addressed data redundancy issue by introducing a low redundancy data based optimal beam prediction approach. In this approach, the authors utilize three steps, namely, feature extraction and fusion, reduction of dimensionality and redundancy by principal component analysis, and prediction with multi-layer perceptron network. In comparison, the authors in \cite{patel2024harnessing} have focused on proposing a multi-user beamforming strategy for mmWave MIMO systems. For that, a multimodal deep neural network model including an alternative loss function, namely supervised soft-contrastive loss, has been proposed to estimate the channel beamspace representations from the sensing data. This estimated beamspaces have been utilized later to identify the radio frequency precoders for all users at the base station, which eventually can help to establish multi-user communication links. 
    
    In another work, Salehi et al. \cite{salehi2024flash} have considered predicting the best sectors specifically in large codebooks, and to achieve this, a multimodal federated learning framework named FLASH-and-Prune has been proposed. In particular, their work starts with predicting the optimal sectors locally by multiple vehicles individually, then aggregates these local model parameters to make a globally optimized model, and finally, a model pruning strategy is applied to further reduce the overheads caused by model parameters exchanging in federated learning settings. In this work, the authors have also prepared a publicly available datasets with real-world measured wireless data, refereed as FLASH\footnote{\url{https://genesys-lab.org/multimodal-fusion-nextg-v2x-communications}} and validated their solution by it.
    
    In addition, the work in \cite{salehi2024multiverse} has designed a digital twin, which closely mimics the real wireless scenarios to speed up the beam selection procedures particularly in unseen environments. In their design, the digital twins are mounted on the vehicles, and collectively coined as multiverse of twins term. The goal of this multiverse of twins is to possibly obtain the necessary ray tracing components and complex interactions at different levels of fidelity from the real world and then utilize them to make decisions on finding the optimum twins based on latency constraints. The results have shown that the proposed Multiverse can offer roughly $85.22$\% for top-$10$ prediction accuracy for unseen scenarios.
    
    To the best of our knowledge, the aforementioned multi-modality related solutions are the most relevant to ours, and accordingly, we show how this current work stands out from existing research in the table \ref{tab: comparison}.

\section{System Model and Problem Statement}
    In this section, we first present the adopted communication system model, and then formally state the multi-modality sensing-aided beam prediction problem addressing in this work.
    
\subsection{System Model}
\begin{figure} [!t]
    \includegraphics[width=\linewidth]{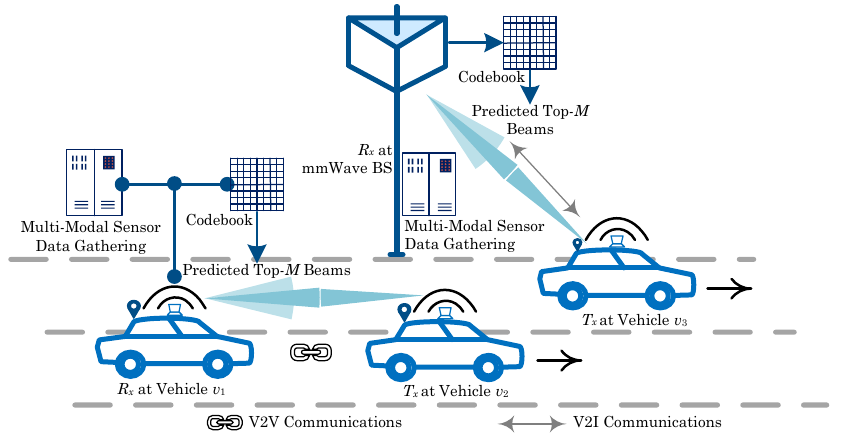}
    \caption{Illustration of our considered mmWave enabled V2I and V2V communications system model.}
    \label{fig: system}
\end{figure}

    In this work, we consider a vehicular communications system model operating at mmWave frequency bands as presented in Fig. \ref{fig: system}. The model consists of a moving vehicle $v_2$ providing communications services, e.g., cooperative perception, to another moving connected vehicle $v_1$ within its coverage area, and these two moving vehicles, $v_1$ and $v_2$, are connected with each other by vehicle-to-vehicle (V2V) communication links. We also consider a stationary base station (typically referred as gNodeB in 5G) unit $\mathcal{A}$ is communicating with the moving connected vehicle $v_3$ to provide communications services, for example, high-definition map updating, within its coverage area by vehicle-to-infrastructure (V2I) communication links. The vehicle $v_1$ and base station $\mathcal{A}$, acting as receivers $R_x$, are equipped with a number of sensing devices, i.e., visual and LiDAR sensors to obtain the multi-modal sensing information from the surroundings, for example, moving vehicles $v_2$ and $v_3$, whereas the vehicles $v_2$ and $v_3$, acting as transmitters $T_x$, have GPS sensing devices to gather the position information, such as latitude and longitude along with the timestamps, and share the information in turn with $v_1$ and $\mathcal{A}$ in real-time, respectively.
    
    Additionally, both the vehicle $v_1$ and base station $\mathcal{A}$ have $N_R$ numbers of antenna array elements, codebooks of $\mathbfcal{Q}$ pre-defined beams with, each elements of codebook referring to a specific beam orientation. The antenna array elements enable $v_1$ and $\mathcal{A}$ to perform beamforming so that $v_1$ and $\mathcal{A}$ can obtain adequate received powers. Each vehicle users $v_2$ and $v_3$ has, for simplicity, one antenna which is oriented always towards the receivers $R_x$. The beam codebooks can be expressed as $\mathbfcal{Q} = \{q_1, q_2, ...., q_{|\mathbfcal{Q}|} : q_i \in \mathbb{C}^{N_R \times 1}\}$, where $|\mathbfcal{Q}|$ is the cardinality of $\mathbfcal{Q}$. The beam vector can be expressed as $(\mathbfcal{Q} \times 1)$.
    
    We further consider that orthogonal frequency-division multiplexing is used in the downlink communications with $\mathcal{N}$ subcarriers. For the given transmitted symbol $s_n \in \mathbb{C}$, the variance of the noise $\sigma^{2}$, and the received complex Additive White Gaussian noise $\omega_n \sim \mathcal{N}_{\mathbb{C}}(0, \sigma_n^{2})$, the downlink transmitted received signal at $n^{th}$ subcarrier at time instant $t$ can be written as:

\begin{equation}
    x_n[t] = \mathbf{h}_n^T[t] q_i[t] s_n + \omega_n[t],
\end{equation}

    where, $\mathbf{h}_n \in \mathbb{C}^{N_R \times 1}$ indicates the channel vectors in each V2V and V2I links. In particular, considering the 3D geometric channel model presented in \cite{va2017inverse}, and denoting $p(\cdot)$ the shaping pulse, $T_s$ the symbol period, $\alpha_l$ the complex path gain, $\tau_l$ the delay, $\textbf{a}_R(\cdot)$ the complex steering vector at receiver end array, $\theta_{l}^A$ elevation angle of arrival, $\phi_{l}^A$ azimuth angle of arrival, of the path $l$, the $\mathbf{h}_n$ can be expressed as frequency-domain at $m^{th}$ symbol vector as follows:
 
\begin{equation}
    \mathbf{h}_n = \sqrt{N_R} \sum_{c=0}^{C-1}\sum_{l=0}^{L-1}\alpha_l e^{-j\frac{2\pi n}{\mathcal{N}}c} p(mT_s - \tau_l)\textbf{a}_R (\theta_{l}^A, \phi_{l}^A),
\end{equation}

    where, $L$ and $C$ denote the number of total channel paths and cyclic prefix length, respectively.

    

\subsection{Problem Statement}
    Given the system model, the primary task of this work is to make predictions for optimal beams $q_i^*$ from the predefined codebooks $\mathbfcal{Q}$ such that the mmWave received powers are maximized. Hence, if the $\mathcal{A}$ and $v_1$ want to determine the optimal beams $q_i^*$ from their beam codebooks $\mathbfcal{Q}$ to serve the $v_3$ and $v_2$, respectively, are determined which maximize the received powers. From the received signal $x_n[t]$ presented in (1), we can get the received power by summing over $\mathcal{N}$ subcarriers at the $\mathcal{A}$ and $v_1$ sides for the receiver codebook $i$. However, if we convert the beam codebooks $\mathbfcal{Q} = \{q_1, q_2, ...., q_{|\mathbfcal{Q}|}\}$ into a unique index as $\mathcal{I} \in \{1, 2, 3, ...., \mathcal{M}: \mathcal{M} \leq |\mathbfcal{Q}|\}$, predicting optimal beams from beam codebooks are same as predicting optimal beam index. Then, predicting the optimal beams problem can be formulated as:
    
\begin{equation}
    \mathcal{I}^* = \underset{i \in \{1, 2, 3, ...., |\mathbfcal{Q}|\}}{\arg\max} \sum_{n=0}^{\mathcal{N}-1} |\mathbf{h}_n^T[t] q_i[t] s_n|^2
\end{equation}

    Specifically, our goal is to predict a set of recommended beams $\mathcal{B} = \{1, 2, 3, ...., M\}$ as top-$M$ beams such that $\mathcal{I}^* \in \mathcal{B}$ as top-1 beam. Nevertheless, to solve the problem in (3), traditional beamforming approaches either leverage the knowledge of $\mathbf{h}_n$, which is rather difficult to acquire in mmWave enabled vehicular scenarios or carry out an exhaustive search over all codebooks thus introducing large training overheads. However, instead of relying on the knowledge of $\mathbf{h}_n$ and an exhaustive search, we aim to utilize deep learning models on the available multi-modal sensory information to solve the optimization problem.

\section{The Proposed Multi-Modal Solution}
    The proposed multi-modal solution is composed of the following three steps: (i) data acquisition, (ii) model training with deep learning, and (iii) deployment in real environment (inference step).
    
\subsection{Data Acquisition}
    We first define the position, visual, and 3D point cloud data matrices at each time instant $t$ as $X_{pos}[t] \in \mathbb{R}^{N_t \times 2}$ (two-dimensional latitude and longitude), $X_{vis}[t] \in \mathbb{R}^{N_t \times w_v \times h_v \times c}$ (width, height, and color channels of visual data), and $X_{lid}[t] \in \mathbb{R}^{N_t \times d_l \times h_l \times w_l}$ (depth, height, and width of point cloud data), respectively, where, $N_t$ denotes the total number of samples for training.
    
    In data acquisition step, the connected vehicle $v_1$ and basestation $\mathcal{A}$ collect the multi-modal sensing data along with corresponding true beams within their own coverage areas in different scenarios (e.g., day and night times), denoted by, $\mathcal{S} = \{s\}_{s=1}^{S}$, and each builds a training dataset $\mathcal{D}_s = \{(X_{pos}^{(s)}, X_{vis}^{(s)}, X_{lid}^{(s)}, Y^{(s)})\}_{s \in \mathcal{S}}$. Here, $Y^{(s)} = \{\mathcal{I}^*_j\}_{j=1}^{|\mathcal{D}_s|}$ represents the corresponding true beams, which can be determined either from the maximum received powers at codesbooks or by utilizing any existing beam selection approaches.

\subsection{Model Training}
    This model training step leverages deep learning architectures to construct a multimodal neural network so that the mapping of input modalities from $\mathcal{D}_s$ to the predicted beam indices can be learned with good generalization abilities. The multimodal network has three unimodal models, which essentially act as feature extractors from the sensing modalities. For that, we define $\psi(\cdot, w)$ as a neural network, where $w$ is its trainable weights which can be optimized to find the optimal weights $w^*$ by solving the following problem:
    
\begin{equation}
    w^* = \underset{w}{\arg\min} \frac{1}{N_t}\sum_{p=0}^{N_t-1} \mathcal{L}(\psi(X^{(p)}, w), Y^{(p)}),
\end{equation}
    
    where, $X \in [X_{pos}, X_{vis}, X_{lid}]$ and $\mathcal{L}(\cdot)$ indicate the loss function to measure how far the neural network's predicted beams deviate from the true beams $Y$. Following that, a softmax function $\mathsf{Softmax}(\cdot)$ is applied on the output of the neural network $\psi(X^{(p)}, w)$ which helps to give the beam prediction result as $\hat{\mathcal{I}} = \mathsf{Softmax}(\psi(X^{(p)}, w))$ by converting the unnormalized output from the neural network into normalized values (a probability distribution). However, since our networks are particularly designed for predicting the top-$M$ beams, the cross-entropy is utilized as a loss function, which can be calculated by $\mathcal{L}(\hat{\mathcal{I}}, Y) = - \frac{1}{N_t}\sum_{p=0}^{N_t-1}\sum_{q=0}^{|\mathbfcal{Q}|-1} Y^{(p)}_{q}\log\hat{\mathcal{I}}^{(p)}_{q}$, where, $|\mathbfcal{Q}|$ is the total number of beam indices, $Y^{(p)}$ is the true beam index for the $p^{th}$ sample, and $\hat{\mathcal{I}}^{(p)}$ is the predicted beam index probability by softmax for $p^{th}$ sample. Further, we utilize a stochastic optimization technique namely adaptive moment estimation (Adam) \cite{KingBa15} for faster convergence on minimizing the loss function while training the neural network models.


\subsection{Inference: Deploying in Real Environment}
    While it may take certain amount of time for data acquisition and model training, after the data acquisition and training, the trained models must have the capability to put into the work in real environment for beams predictions with new data. This step is often referred as inference step. Consequently, the trained models is utilized in the inference step to make probabilistic predictions of top-$M$ beams within the coverage areas as 

\begin{equation}
    \hat{\mathcal{I}} = \mathsf{Softmax}(\psi_{w^*}(X_{pos}^{n}, X_{vis}^{n}, X_{lid}^{n})),
\end{equation}
    
    where, $\psi_{w^*}$ is parameterized by optimal weights $w^*$, and $X_{pos}^{n}$, $X_{vis}^{n}$, and $X_{lid}^{n}$ denote the new input samples from GPS, Visual, and LiDAR sensors, respectively.

    However, when it comes to real environment deployment, certain prediction errors may arise leading to performance degradation. To address this, the trained model may need to update with relatively smaller datasets than the original ones and fine-tune in accordance with the performance requirements.
    
\section{Feature Extractions on Multi-Modal Sensing}
    In this section, the proposed feature extractions from multi-modal sensory data inputs is described in detail highlighting the data processing steps, the architecture of feature extractions, and finally, beam selection procedures.

\subsection{Data Preprocessing}
    According to Section IV, the training dataset is required to feed into multimodal neural networks as inputs, and the inputs are desired to be fixed and consistent in size. In this regard, feeding multimodal sensing data directly may require developing deep learning models with high complexity in terms of both architecture and computational cost. Thus, to make it fit as well as accelerate the convergence of the deep learning models, we process the sensing data into following preprocessing procedures.
    
    \textit{Position Data Preprocessing:} The raw position data captured from GPS sensors are basically in Decimal Degrees, specifically the latitude and longitude values are from $-90^{\circ}$ to $+90^{\circ}$ and $-180^{\circ}$ to $+180^{\circ}$, respectively. Hence, we pass through the raw position data into a preprocessing procedure, namely data normalization. For that, we calculate the normalized values of latitude $x_{lt}^{\prime}$ and longitude $y_{lg}^{\prime}$ at each rows of the matrix  as: $(x_{lt}^{\prime}, y_{lg}^{\prime}) = ((x_{lt} - x_{lt\_min})/(x_{lt\_max}) - x_{lt\_min}), (y_{lg} - y_{lg\_min})/(y_{lg\_max} - y_{lg\_min}))$, within Cartesian coordinate system, where $x_{lt}$ and $y_{lg}$ are the raw latitude and longitude values, while the $max$ and $min$ denote their maximum and minimum values, respectively.

    \textit{Visual Data Preprocessing:} Converting the obtained visual image data into the inputs for the expected model involves two procedures, i.e., resizing and normalization. In resizing, we fix the spatial dimensions of the visual images into specific $224 \times 224 \times 3$ shapes, where $224 \leq w_v$ and $224 \leq h_v$. Here, $224 \times 224$ indicates that each images has a resolution of $224$ pixels in width and $224$ pixels in height, for a total of $224 \times 224 = 49,536$ pixels. And, 3 represents the red, green, and blue color channels range from $0$ to $255$, which undergo a normalization by subtracting the mean values $(0.485, 0.456, 0.406)$ from them and then, dividing the results by the standard deviation values $(0.229, 0.224, 0.225)$ so that the normalized values become zero mean and unit variance.
    
    \textit{Point Cloud Data Preprocessing:} For each sample, as a result of raw measurements by LiDAR sensors output a large point cloud data $\boldsymbol{P} = \bigcup_{i=0}^{\boldsymbol{n}-1} \{(x_i, y_i, z_i)\}$, where $\boldsymbol{n}$ is the point numbers, and $\{(x_i, y_i, z_i)\}$ denotes the coordinate of the $i^{th}$ point representing the obstacle. Even though our expected model will be able to handle the raw point clouds directly, the point clouds samples may have varying number of points. Hence, we preprocess the $\boldsymbol{P}$ either by padding or downsampling while preserving the underlying essential features so that each sample will have consistency in terms of fixed number of points, and we set the fixed number of points as $15,000$. In padding, we pad the rest of points with zeros if the number of points in samples have lesser than the desired points whereas, in downsampling, we randomly eliminate the points otherwise.
    
    Finally, the revised matrices with preprocessed data at time stamp $t$ are then be denoted as $X^{\prime}_{pos}[t]$, $X^{\prime}_{vis}[t]$, and $X^{\prime}_{lid}[t]$, respectively.
    
\subsection{Multi-Modal Feature Extractions Architecture}
\begin{figure*} [!t]
	\includegraphics[width=.95\linewidth]{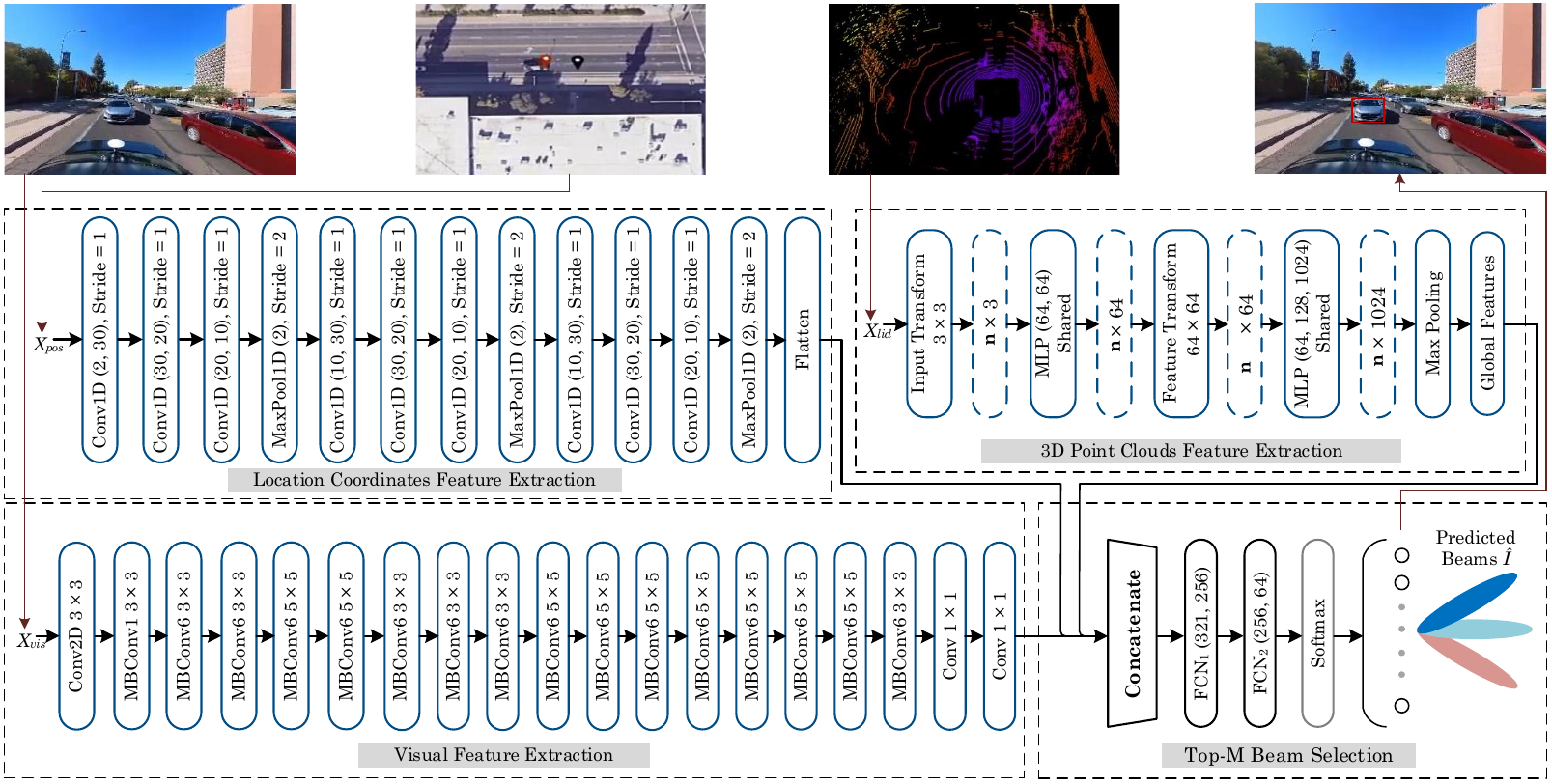}
    \centering
    \caption{The details diagram of the proposed deep learning model on multi-modality sensing for mmWave beamforming, which is mainly composed of three feature extraction and one top-$M$ beam selection components.}
    \label{fig: DL-Model}
\end{figure*}

    The designed deep learning architecture of proposed feature extractions for mmWave beamforming task is shown in Fig. \ref{fig: DL-Model}. In this architecture, we basically choose a type of deep learning techniques namely, convolutional neural networks. Due to having convolutional operations, such neural networks can achieve very good performance on visual data, however, we also utilize the same in other two modalities so that the resulting corresponding features can be fused in a straightforward way. In the following, we present the detailed feature extractions from each modalities.

    \textit{1) Position Feature Extraction:} The proposed feature extraction model for position data adopts two components: the convolutional blocks and a flatten layer. Specifically, the convolutional blocks, the key building blocks used in this proposed model, are made up of three convolutional blocks. These convolutional blocks are formed together with three sequence of 1D (one-dimensional) convolutional layers, however, each blocks are followed by a max-pooling layer.
    
    Given the preprocessed position data $X^{\prime}_{pos}$ as an input to the proposed model, the convolutional block $\mathsf{ConvNetBlock}_1$ first compute the inputs to extract the features by learning the hidden meaningful representations. This block is repeated twice as $\mathsf{ConvNetBlock}_2$ and $\mathsf{ConvNetBlock}_3$ in the network to make enhanced feature extraction and improved performance. However, the max-pooling layers in all blocks contribute to downsize the feature maps while keeping the most significant features, leading to decreasing the risk of the model becoming overfitted. Beyond the convolutional blocks, the output from $\mathsf{ConvNetBlock}_3$ is passed though the flatten layer, which is employed to flatten the output received from the last convolutional block, while not effecting the batch size. The overall process is presented as follows.

\begin{equation}
    \begin{array}{l}
    c_1 = \mathsf{ConvNetBlock}_1(g_t) \\
    c_2 = \mathsf{ConvNetBlock}_2(c_1) \\
    c_3 = \mathsf{ConvNetBlock}_3(c_2),
    \end{array}
\end{equation}

    where, $c_1$, $c_2$, and $c_3$ are the hidden vectors, and $g_t$ is considered as a variable representing the vector elements of $X^{\prime}_{pos}[t]$.

    \textit{2) Visual Feature Extraction:} For the extraction of visual data, we employ fine-tuned EfficientNet-B0 model \cite{tan2019efficientnet}. This model utilizes a compound scaling method, which applies an uniform scaling over depth, width, and resolution all together to the baseline model depending on the available resources. In particular, the baseline model of EfficientNet-B0 comprises two types of building blocks, namely, the 2D convolutional (Conv2D) and mobile inverted bottleneck convolution (MBConv) blocks. With these building blocks, the model works in the following ways.
    
    At first, the Conv2D block performs an initial convolutional operation by applying $3 \times 3$ filter on the preprocessed visual data input $X^{\prime}_{vis}$ to capture the simple basic features, such as edges and textures. Besides, this operation also utilizes a stride of $2$, which basically downsamples the feature maps with reducing the spatial dimensions, i.e., heights and widths, of the inputs by a factor of $2$ so that the subsequent blocks will have ability to contribute on recognizing more complex and abstract features.

    The resulting outputs are passed through a number of MBConv blocks. The MBConv blocks are designed by inverter residual block structure and utilizes depthwise separable convolution along with an additional squeeze and excitation layer. The depthwise separable convolutional process deals with both spatial dimensions (height and width) and depth dimensions (number of channels) by following depthwise convolution and pointwise convolution operations after one another.

    Based on expanding the number of channels in expansion phase, the MBConv blocks have two variants including MBConv1 (no expansion) and MBConv6 (expansion by a factor of $6$). From a structural point of view, each MBConv blocks start with expanding the channel inputs by a $1\text{×}1$ convolution, which then utilizes $3\text{×}3$ or $5\text{×}5$ depthwise convolution, and finally a $1\text{×}1$ pointwise convolution to adjust the dimensionality of the outputs so that the inputs and outputs can be added through the skip connection. The above processes can be represented as follows.

\begin{equation}
    \begin{array}{l}
    \text{Conv}(x) = \mathsf{Swish}(\mathsf{BNorm}(\mathsf{Conv}1\text{×}1(x))) \\
    \text{DConv}(y) = \mathsf{SE}(\mathsf{SiLU}(\mathsf{BNorm}(\mathsf{DConv}(\text{Conv}(x))))) \\
    \text{PConv}(z) = \mathsf{Swish}(\mathsf{BNorm}(\mathsf{PConv}1\text{×}1(\text{DConv}(y)))),
    \end{array}
\end{equation}

    where, $\mathsf{Conv}1\text{×}1(\cdot)$, $\mathsf{DConv}(\cdot)$, $\mathsf{PConv}(\cdot)$, $\mathsf{BNorm}(\cdot)$, and $\mathsf{SiLU}(\cdot)$ are the functions of $1\text{×}1$ convolution, depthwise convolution, pointwise convolution, batch normalization, and activation, respectively. 

    Following the MBConv blocks, the outputs undergo a final convolutional block which involves a $1\text{×}1$ convolution to aggregate the features learned from previous blocks and yield a large number of channels ($1280$ in our case), and then, a global average pooling to make an average of each feature map obtained from the final convolutional block, resulting in reducing the spatial dimensions to a single value. And, in this way, the detailed features from visual inputs can be captured. Besides, it should be noted that the batch normalization and SiLU activation are applied after each convolutional operations. Here, the Sigmoid Linear Unit (SiLU), also known as Swish, function is a type of smooth and non-linear activation function, and it can be defined as:

\begin{equation}
    \mathsf{SiLU}(x) = x \times \mathsf{Sigmoid}(x), 
\end{equation}
    
    where, $\mathsf{Sigmoid}(x) = (1 + \exp(-x))^{-1}$ denotes as the sigmoid function.

    \textit{3) Point Cloud Feature Extraction:} Motivated by the capabilities of handling unordered 3D point clouds directly as well as having invariance properties under order and transformations, we utilize fine-tuned PointNet model \cite{qi2017pointnet}. In particular, the PointNet model consists of three parts: (a) input and feature transformations, (b) shared multi-layer perceptrons (1D convolution), and (c) max pooling. The model works in the following three ways:

    The primary role of input transformation is to take point cloud inputs from $X^{\prime}_{lid}$, and learns a $3\text{×}3$ transformation matrix which is then multiplied to every input points to obtain invariance property to transformation, such as rotation and translation by aligning the input point clouds in a canonical coordinate space. In input points shape $\boldsymbol{n}\text{×}3$, $\boldsymbol{n}$ represents the total number of points in each point cloud samples, whereas $3$ refers the $X$, $Y$, and $Z$ coordinates. After applying input transformation, the transformed data ($3$ dimensions) undergoes a series of multi-layer perceptrons $(64, 64)$ to learn the individual local features through processing each point independently, which returns an output with uplifting the dimensions to $64$. While capturing such local features, the same weights are shared for each point, thereby achieving invariant to the order of the points.

    Next, the feature transformation takes the individual point features obtained in first part as inputs $(\boldsymbol{n}\text{×}64)$. Similar to the input transformation, it repeats to learn a matrix again, which is then applied across all individual point features for further alignment and refining on learning the local features, allowing the outputs have invariance under different scales. Instead of $3\text{×}3$, now the transformation matrix becomes $64\text{×}64$. After feature transformation, let $\boldsymbol{P} = \{p_1, p_2, ...., p_{\boldsymbol{n}}\}$ as the outputs, where, $\boldsymbol{p}_i$ can be the $i^{th}$ point within the point clouds. These outputs are fed later into the an another shared multi-layer perceptrons $(64, 128, 1024)$ to emphasize on obtaining features for subsequent processing. Here, the $64, 128, 1024$ refers to the inputs mapping start with dimensions of $64$, to $128$, and at the end, to $1028$. This can be denoted as $f_i = \mathsf{MLP}(\boldsymbol{p}_i)$, where $\mathsf{MLP}$ indicates multi-layer perceptrons, and $f_i$ is the local feature of $i_{th}$ point $\boldsymbol{p}_i$. Particularly, the multi-layer perception is comprised of a number of fully connected layers along with a non-linear activation function namely ReLU (Rectified Linear Unit). For all features, we can write as follows:
    
\begin{equation}
    \begin{array}{l}
    f_1 = \mathsf{ReLU}(W \times p_1 + \mathsf{bias}) \\
    f_2 = \mathsf{ReLU}(W \times p_2 + \mathsf{bias}) \\
    f_3 = \mathsf{ReLU}(W \times p_3 + \mathsf{bias}) \\
    . \\
    . \\
    . \\
    f_{\boldsymbol{n}} = \mathsf{ReLU}(W \times {p}_{\boldsymbol{n}} + \mathsf{bias}),
    \end{array}
\end{equation}

    where, $W$ and $\mathsf{bias}$ are the corresponding weights and biases, respectively.

\begin{figure*} [!t]
\begin{subfigure}[b]{\textwidth}
	\centering
	\includegraphics[width=9.6cm]{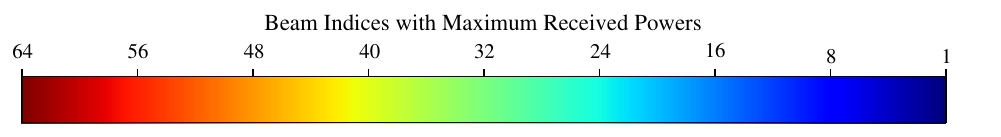}
\end{subfigure}
\begin{subfigure}[b]{0.24\textwidth}
    \centering
    \includegraphics[width=4.4cm]{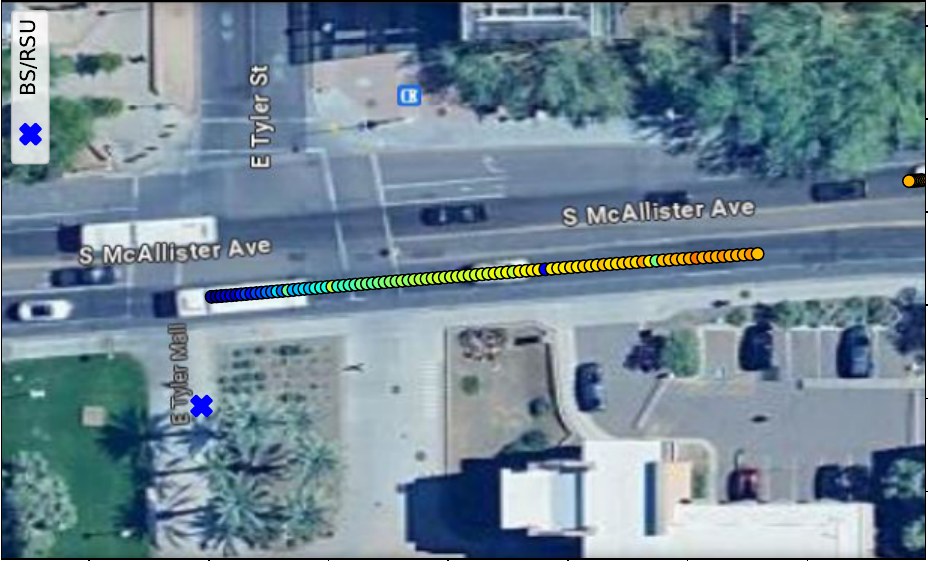}
	\caption{\centering \scriptsize For scenario 31}
\end{subfigure}
\begin{subfigure}[b]{0.24\textwidth}
	\centering
	\includegraphics[width=4.4cm]{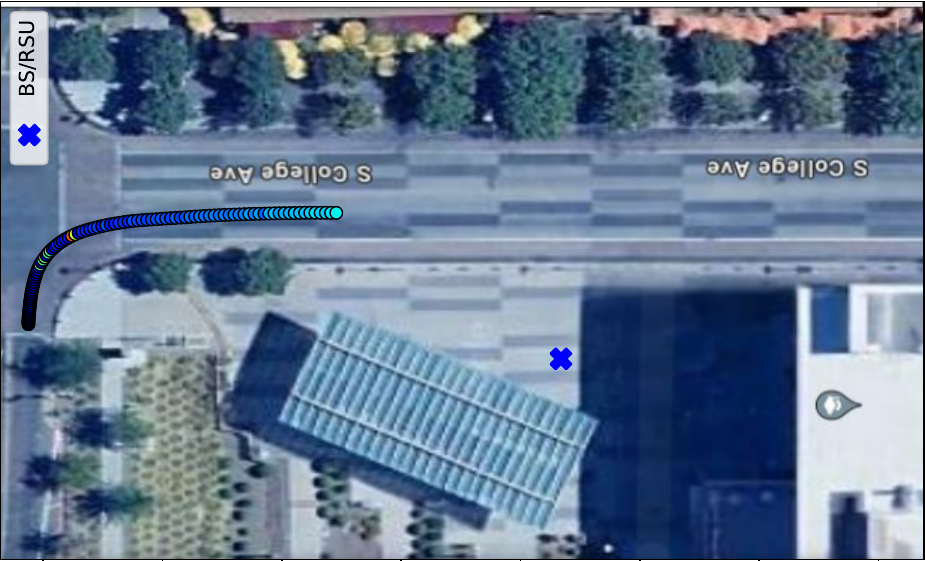}
	\caption{\centering \scriptsize For scenario 32}
\end{subfigure}
\begin{subfigure}[b]{0.24\textwidth}
    \centering
    \includegraphics[width=4.4cm]{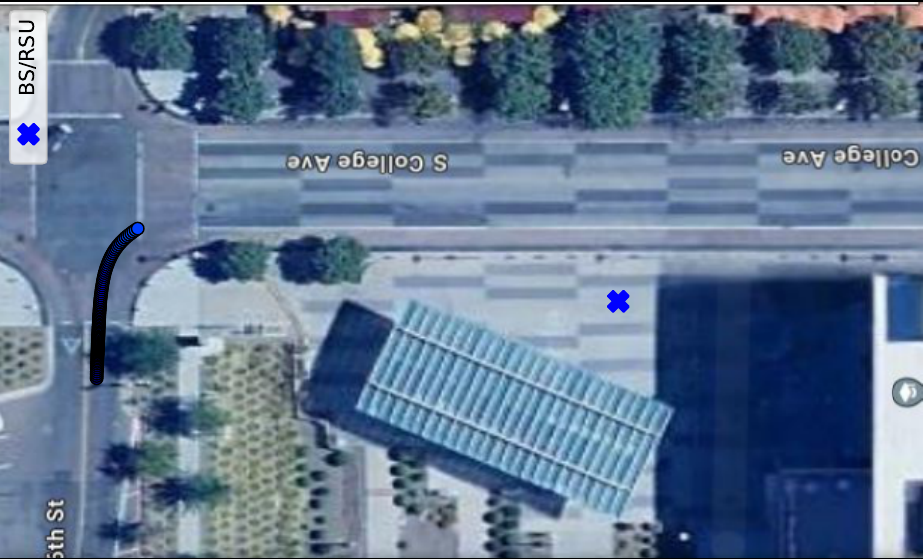}
    \caption{\centering \scriptsize For scenario 33}
\end{subfigure}
\begin{subfigure}[b]{0.24\textwidth}
    \centering
    \includegraphics[width=4.4cm]{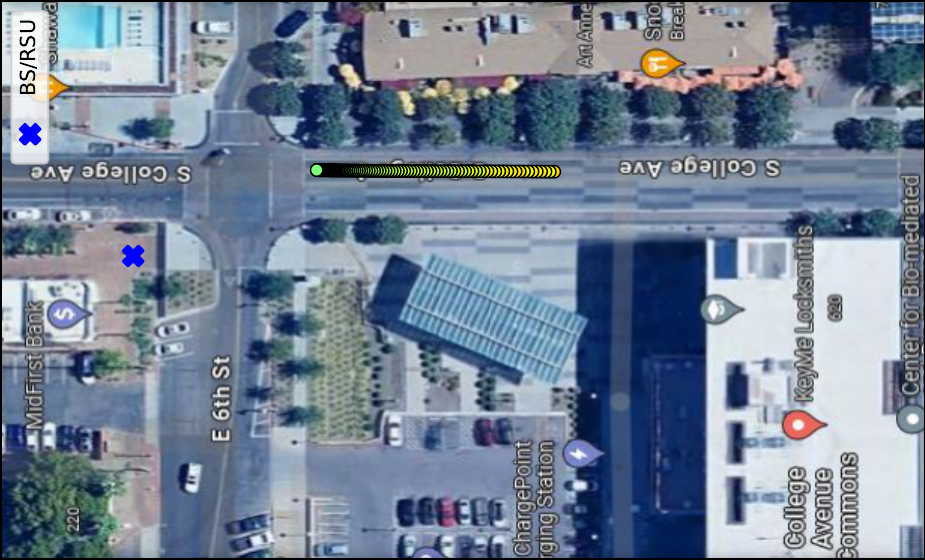}
    \caption{\centering \scriptsize For scenario 34}
\end{subfigure}

\vspace{3mm}

\begin{subfigure}[b]{0.24\textwidth}
    \centering
    \includegraphics[width=4.4cm]{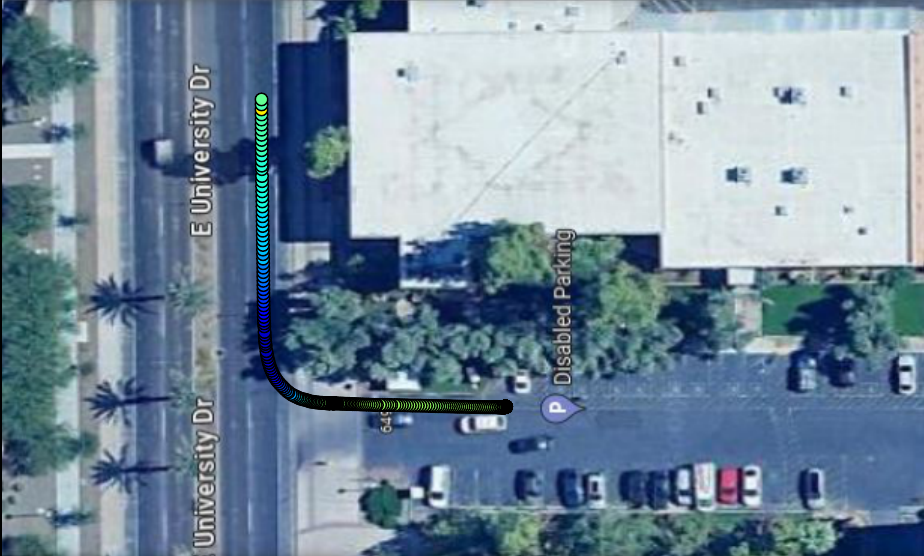}
	\caption{\centering \scriptsize For scenario 36}
\end{subfigure}
\begin{subfigure}[b]{0.24\textwidth}
	\centering
	\includegraphics[width=4.4cm]{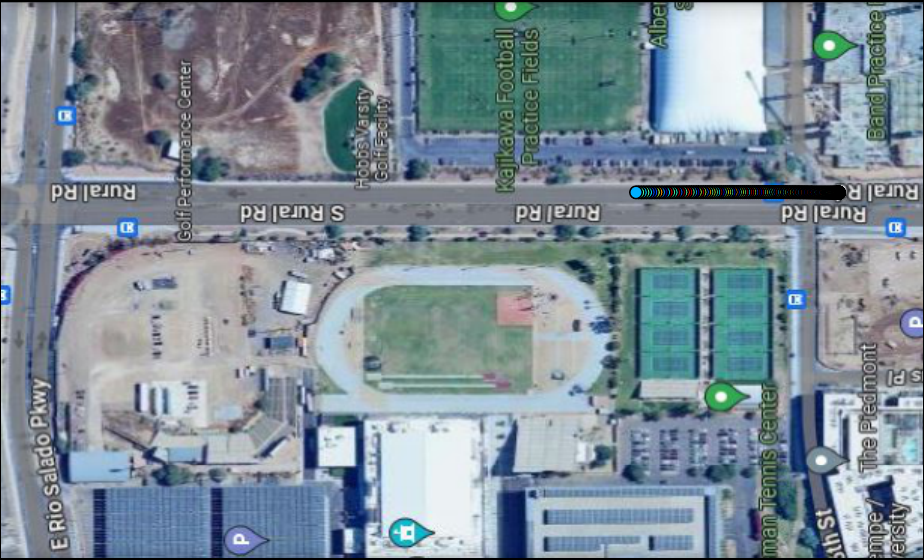}
	\caption{\centering \scriptsize For scenario 37}
\end{subfigure}
\begin{subfigure}[b]{0.24\textwidth}
    \centering
    \includegraphics[width=4.4cm]{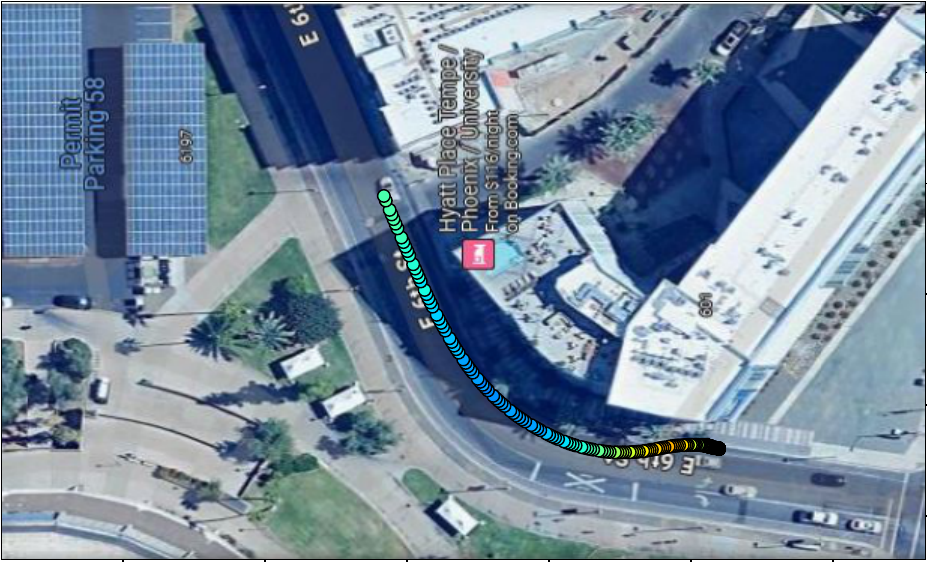}
    \caption{\centering \scriptsize For scenario 38}
\end{subfigure}
\begin{subfigure}[b]{0.24\textwidth}
    \centering
    \includegraphics[width=4.4cm]{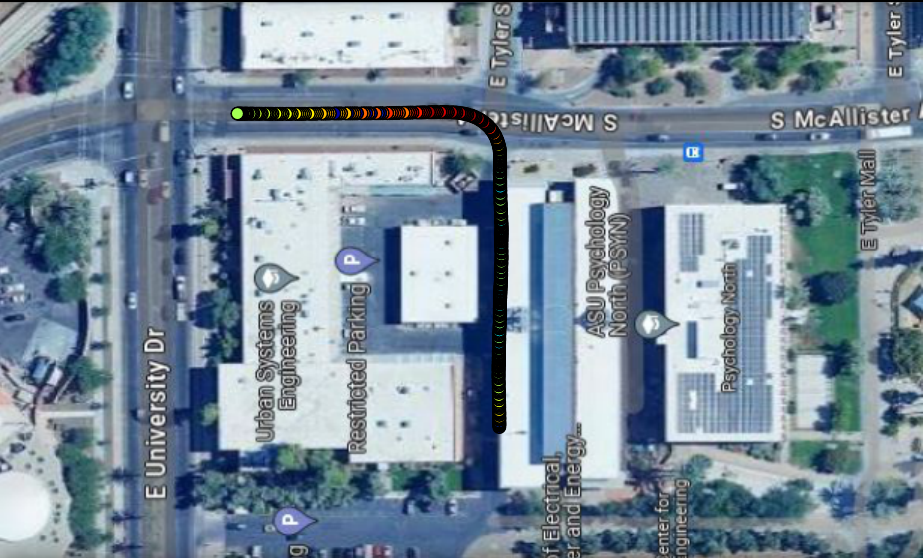}
    \caption{\centering \scriptsize For scenario 39}
\end{subfigure}
    \caption{Visual representation of receiver vehicle’s GPS location data points (100 and 400 samples for V2I and V2V scenarios, respectively) along with corresponding best beam indices out of 64 beams on Google Map satellite view.}
	\label{fig: Maps}
\end{figure*}
    
    Serving as the last operation in max pooling, it aggregates all individual point features, such as $f_1, f_2, f_3, ...., f_{\boldsymbol{n}}$ obtained from last shared multi-layer perceptrons, which finally results in a constructed global feature representation, which can be denoted as

\begin{equation}
    \mathcal{G} = \mathsf{MaxPool}(f_1, f_2, ...., f_{\boldsymbol{n}}), 
\end{equation}

    where, $\mathsf{MaxPool}$ is a max pooling function adhering a symmetric function operation, thereby making the model to work regardless of point orders.
    
\subsection{Top-$M$ Beams Selection}
    Assuming the outputs from each feature extractors as $\mathsf{z}_{pos}$, $\mathsf{z}_{vis}$, and $\mathsf{z}_{lid}$, receptively, the beam selection module takes them as inputs. These output features $\mathsf{z}_{pos}$, $\mathsf{z}_{vis}$, and $\mathsf{z}_{lid}$ are then fused along the feature dimensions, and we utilize concatenation to fuse the features. The resulting concatenated features, after that, is forwarded to the two fully connected (dense) layers $\mathsf{FCN}_1$ and $\mathsf{FCN}_2$. The two fully connected layers together encompass a prediction by applying weights and biases. The weights and biases of these layers will be learned during the training process, enabling the multi-modal model to map input features to the desired output. We can describe as
    
\begin{equation}
    \begin{array}{l}
    \chi = \mathsf{Concate}(\mathsf{z}_{pos}, \mathsf{z}_{vis}, \mathsf{z}_{lid}) \\
    \mathcal{C}_1 = \mathsf{FCN}_1(\chi) \\
    \mathcal{C}_2 = \mathsf{FCN}_2(\mathcal{C}_1),
    \end{array}
\end{equation}

    where, $\mathsf{Concate}$ is the concatenation function, and 
    $\chi$, $\mathcal{C}_1$ and $\mathcal{C}_2$ denote the results from concatenation operation and the two connected layers, respectively.
    
    In the end, the softmax layer is involved as last layer, where a softmax activation function ($\mathsf{Softmax}(\cdot)$) is applied on the output from the second fully connected layer to transform the values into probability scores within the range $[0, 1]$ with total value 1, which can be interpreted to eventually determine the top-$M$ beams $\hat{\mathcal{I}} = \mathsf{Softmax}(\mathcal{C}_2)$.

\section{Experiments and Evaluations}
    In this section, we present the datasets description, implementation settings, and the results obtained from a series of experiments. Subsequently, we also present the performance comparison with existing standard defined approach in terms of computational and communications overheads.

\subsection{Datasets Description}
    In this work, we consider the DeepSense 6G dataset \cite{alkhateeb2023deepsense, morais2024deepsense}, a collection of real-world mmWave sensing and communications measurements, to evaluate the effectiveness of proposed top-$M$ beam prediction introduced in Section V. Specifically, consistent with the considered system model and formulated problem, we adopt scenario $31$ ($7,012$ samples), scenario $32$ ($3,235$ samples), scenario $33$ ($3,981$ samples), scenario $34$ ($4,439$ samples), scenario $36$ ($24,800$ samples), and $37$ ($31,000$ samples), $38$ ($36,000$ samples), and $39$ ($20,400$ samples). In particular, the scenarios $31$ to $34$ are for V2I communications, whereas the scenarios $36$ to $39$ are for V2V communications. These scenarios are captured with two testbed setups including the stationary unit ($\mathcal{A}$) along with one moving vehicle ($v_3$), and two moving vehicles (namely, $v_1$ and $v_2$) deployed in diverse outdoor urban scenarios having long and shorter distances. And, the data has been collected at Tempe, Phoenix, and Scottsdale of Arizona during both day and night. Most importantly, this dataset has the mapping between beam codebooks and measured corresponding received powers for line-of-sight (LoS) conditions in both V2I and V2V scenarios. Other related datasets on multi-modal sensing and communications, such as Raymobtime \cite{raymobtime} and FLASH \cite{flash} do not consider V2V scenarios, which limit their use in this work. However, it should be noted that since our proposed solution already utilizes multi-modal data, which could be further fine-tuned and revised to adjust beamforming prediction in non-line-of-sight (NLoS) scenarios, such as including alternative relay node selection task. 

    
    Here, the unit $\mathcal{A}$ and unit $v_1$ act as receivers $R_x$, whereas the other units $v_3$ and $v_2$ act as transmitters $T_x$, and both are operating at $60$ GHz carrier frequencies. Additionally, the units $\mathcal{A}$ and $v_1$ employ mmWave phased arrays where, each of the phased arrays has $16$ elements ($N_R = 16$) uniform linear arrays (ULA). The unit $v_1$, in particular, has four mmWave phased arrays which are facing right, left, back, and front directions on the vehicle. The number of pre-defined and over-sampled codebook beams for both phased array of units $\mathcal{A}$ and $v_1$ are $64$ ($|\mathbfcal{Q}| = 64$), whereas the units $v_3$ and $v_2$ have one antenna element each with a total $64 \times 1$ = $64$ beam pairs.

\begin{figure*} [!t]
\begin{subfigure}[b]{\textwidth}
    \centering
	\includegraphics[width=0.24\textwidth]{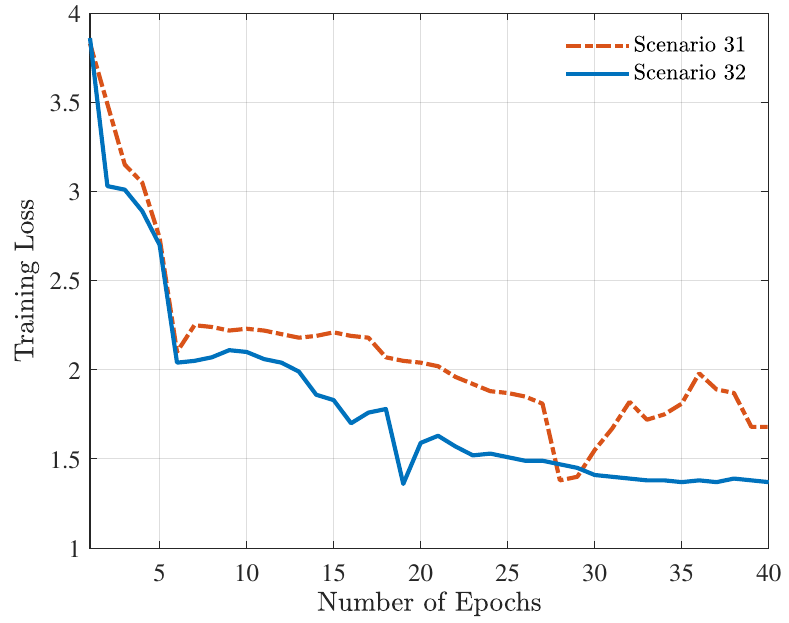}
    \includegraphics[width=0.24\textwidth]{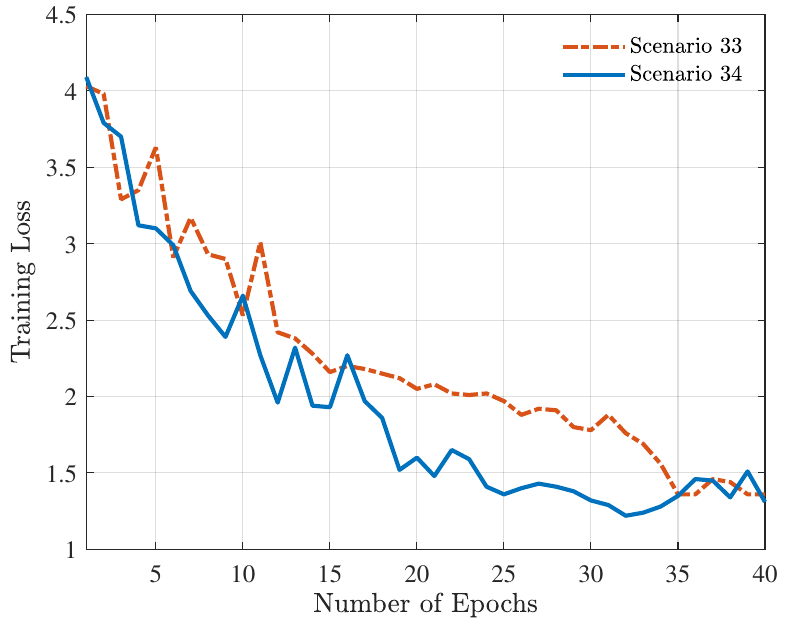}
    \includegraphics[width=0.24\textwidth]{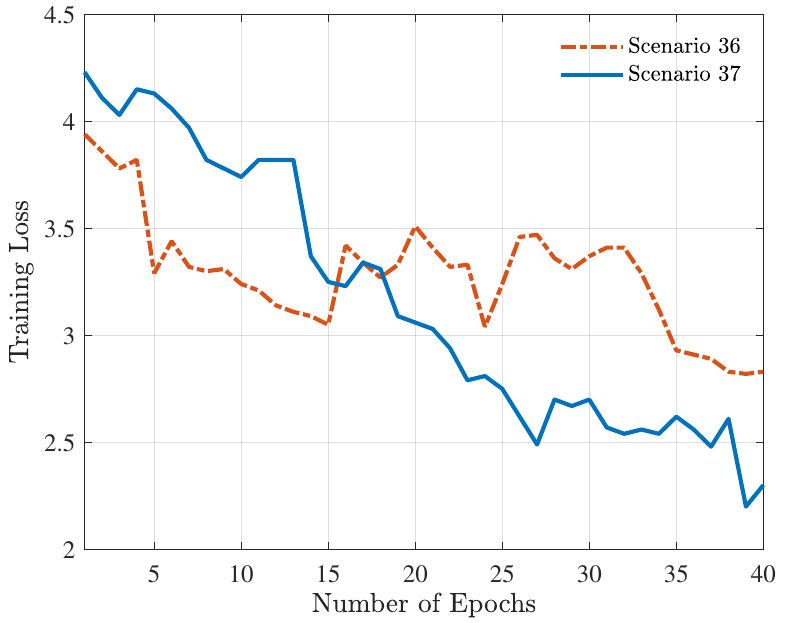}
    \includegraphics[width=0.24\textwidth]{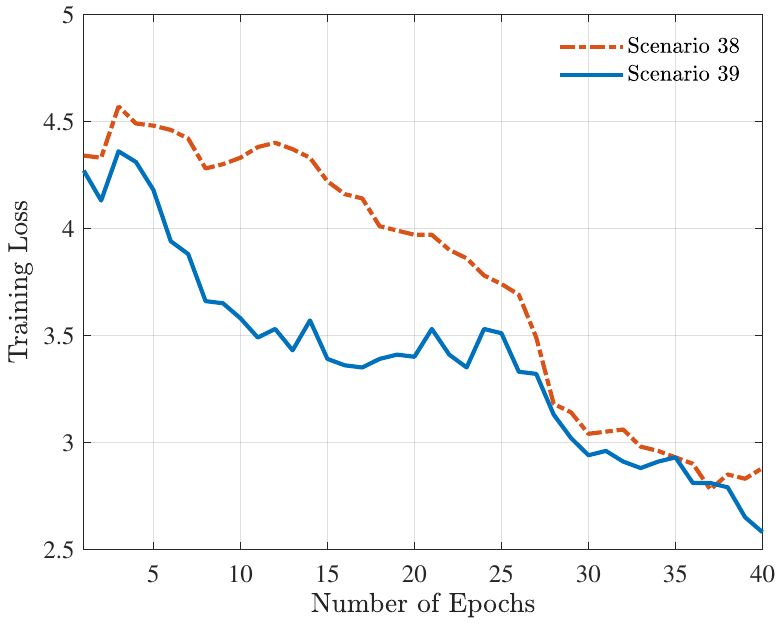}
	\caption{\centering \scriptsize The loss curve on the training data from all considered scenarios measuring the difference between predicted and true beams.}
\end{subfigure}

\vspace{3mm}

\begin{subfigure}[b]{\textwidth}
    \centering
	\includegraphics[width=0.24\textwidth]{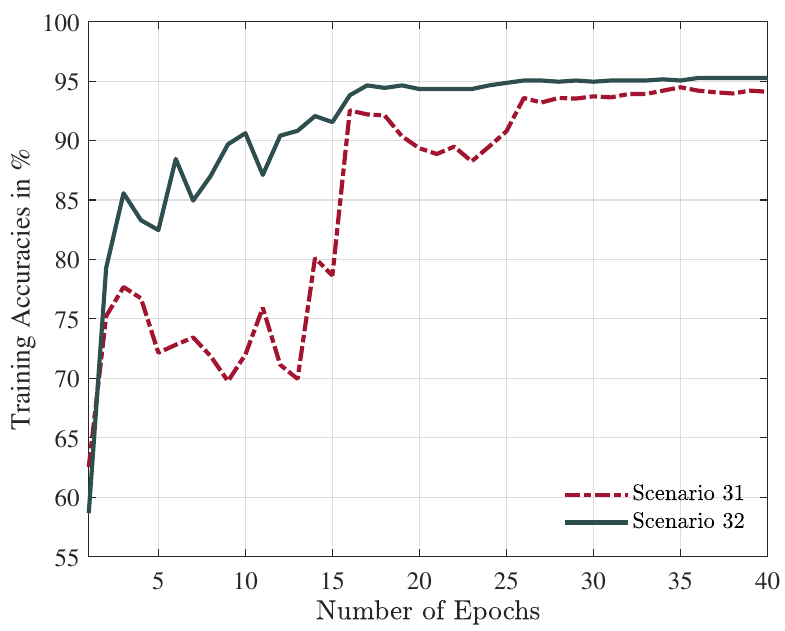}
    \includegraphics[width=0.24\textwidth]{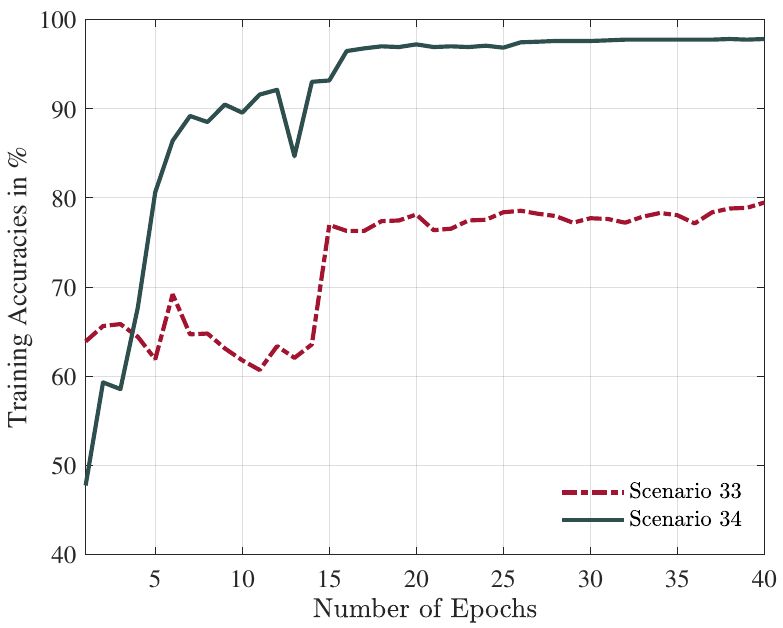}
    \includegraphics[width=0.24\textwidth]{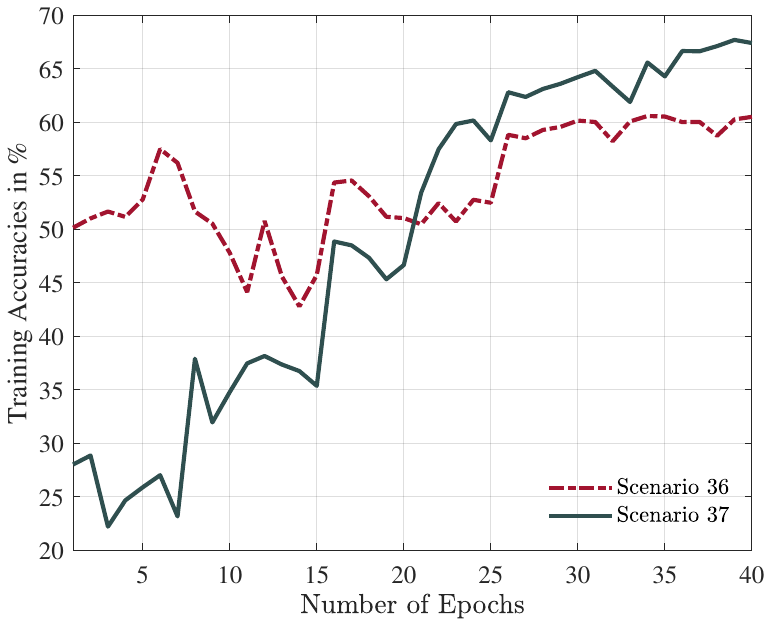}
    \includegraphics[width=0.24\textwidth]{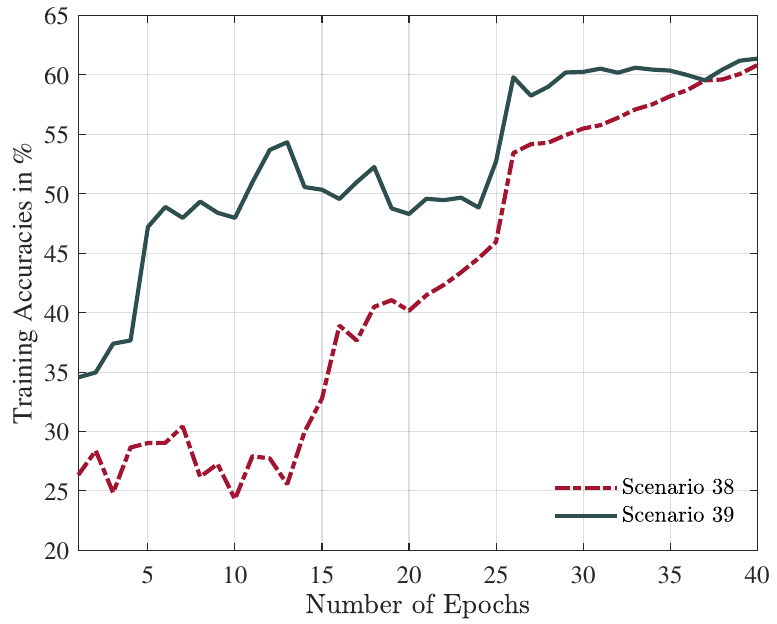}
	\caption{\centering \scriptsize The training accuracies obtained at each epochs representing the model performance improves over epochs.}
\end{subfigure}
    \caption{The results of loss and accuracies while training the proposed model, indicating how the learning performances of the model on training data from all considered scenarios are improving over $40$ number of epochs.}
	\label{fig: loss_training_accuracies}
\end{figure*}
    
    In particular, each of the units $v_3$ and $v_2$ utilizing its single antenna element continually performs omni-directional simultaneously, the units $\mathcal{A}$ and $v_1$ collect received powers via a full beam sweeping. The received powers at beams are represented as beam indices ($\mathcal{I} \in \{1, 2, 3, ...., 64\}$). Each of the units $\mathcal{A}$, $v_1$, $v_2$, and $v_2$ is also equipped with GPS Real Time Kinematics sensors to obtain the vehicle position, i.e., latitude and longitude values in real-time. The time interval between two consecutive samples is $0.1$ second corresponding to a rate of $10$ samples/second. At each instant, the units $\mathcal{A}$ and $v_1$ record the synchronous data of location coordinates of units $v_3$ and $v_2$ as well as corresponding received powers and optimal beam indices (referred as ground truth indices). Figs. \ref{fig: Maps}(a)-(h) illustrate the visual representation of $100$ and $400$ location data points for V2I and V2V scenarios, respectively of $R_x$ corresponding to individual ground truth beams associated with maximum received powers for all scenarios. In fact, this indicates how the best beams change as the vehicles' change their respective position.

\subsection{Implementations and Performance Evaluation}
    A series of experiments are carried out to assess the performance of the designed multi-modal model on the considered real-world datasets. The experiments are conducted on a computing device having Intel Core i7-10875H CPU with 32 GB RAM and NVIDIA GeForce RTX 2080 Super GPU with 8 GB memory. In particular, the deep learning model is developed by PyTorch and CUDA toolkit 11.7. We split the data resources of each scenario into $60$\%, $20$\%, and $20$\% respectively for training, validation, and testing. For the training step, the fixed learning rate and weight decay values of the Adam optimizer are set to $0.01$ and $1 \times 10^{-4}$, respectively, while the batch size is $16$. With these parameters, we then train the model on splitted training samples of each individual scenarios, and the model is trained up to a total of $40$ epochs. Figs. \ref{fig: loss_training_accuracies}(a) and (b) show the results of the training loss and accuracies, respectively, obtained at each epoch. The data preprocessing and implementation codes along with detailed instructions are publicly available to facilitate the reproducibility of this work at GitHub\footnote{\url{https://github.com/mbaqer/V2X-mmWave-Beamforming}}.   

\begin{figure*} [!t]
\begin{subfigure}[b]{\textwidth}
    \centering
	\includegraphics[width=0.24\textwidth]{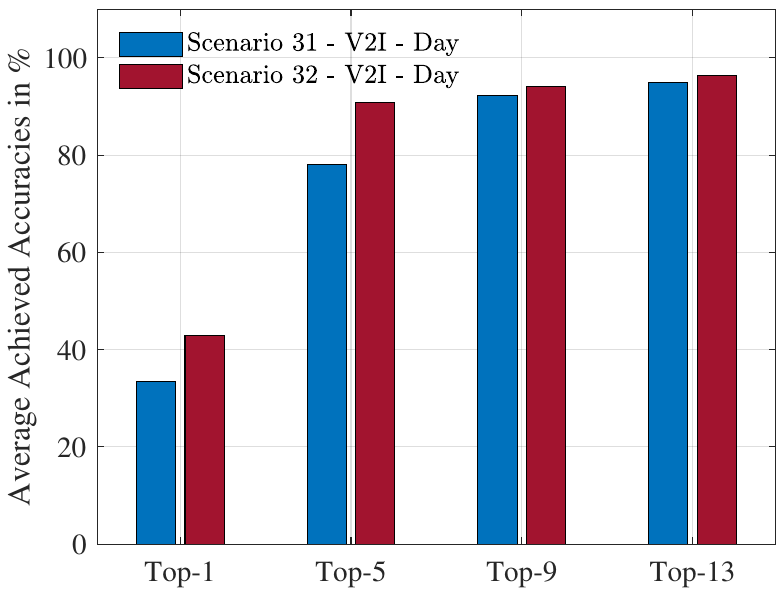}
    \includegraphics[width=0.24\textwidth]{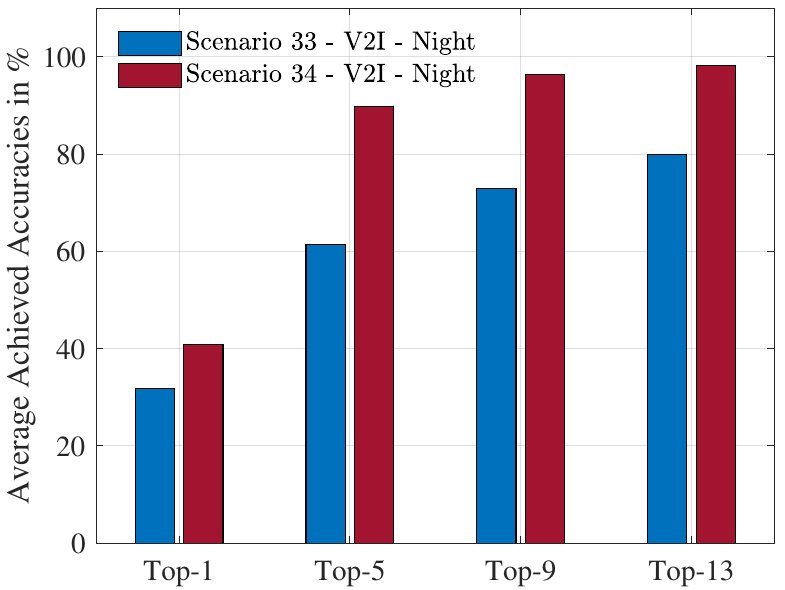}
    \includegraphics[width=0.24\textwidth]{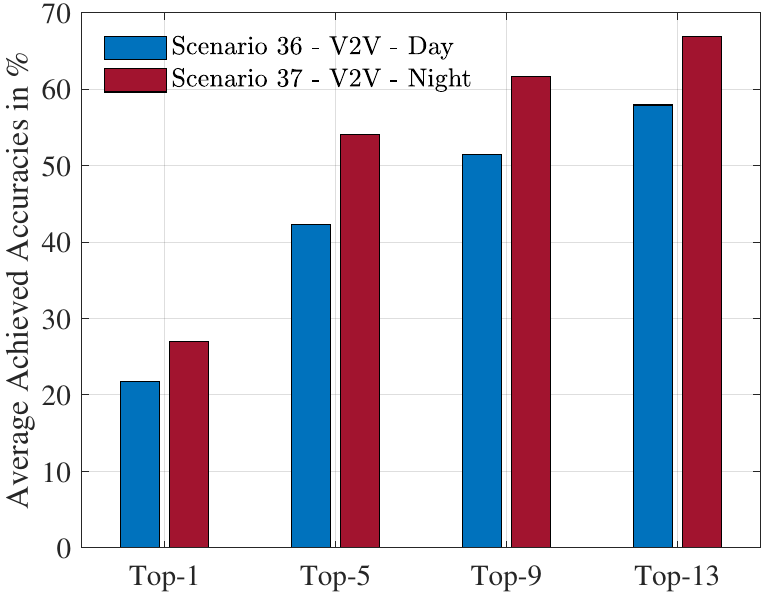}
    \includegraphics[width=0.24\textwidth]{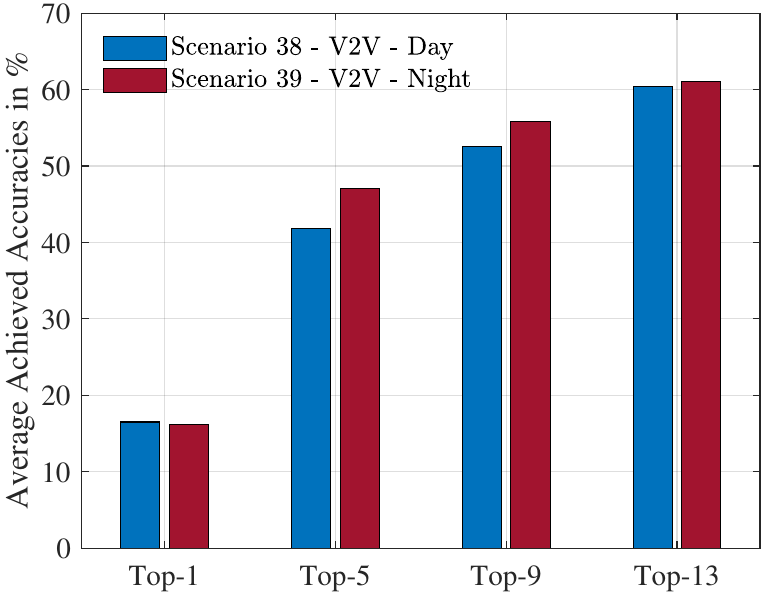}
	\caption{\centering \scriptsize Average achieved accuracies on the testing data considering all three sensing modalities}
\end{subfigure}

\vspace{3mm}

\begin{subfigure}[b]{\textwidth}
    \centering
	\includegraphics[width=0.24\textwidth]{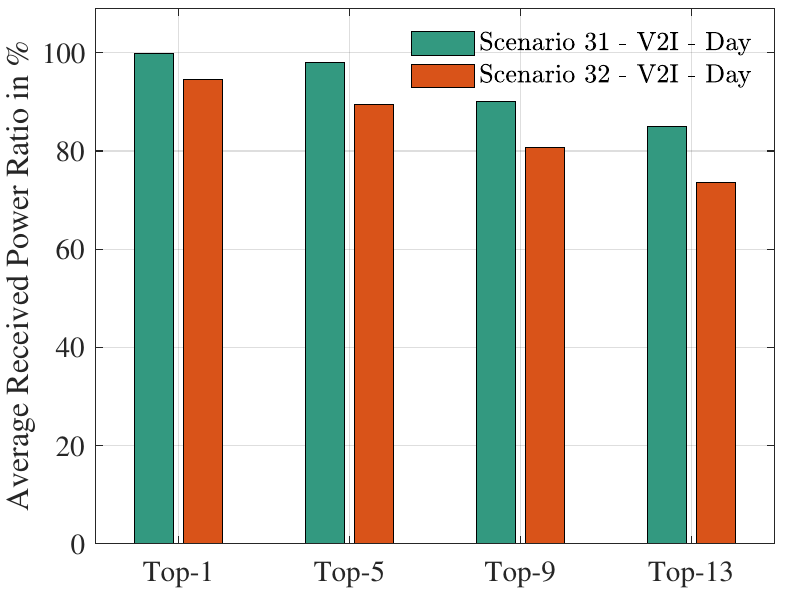}
    \includegraphics[width=0.24\textwidth]{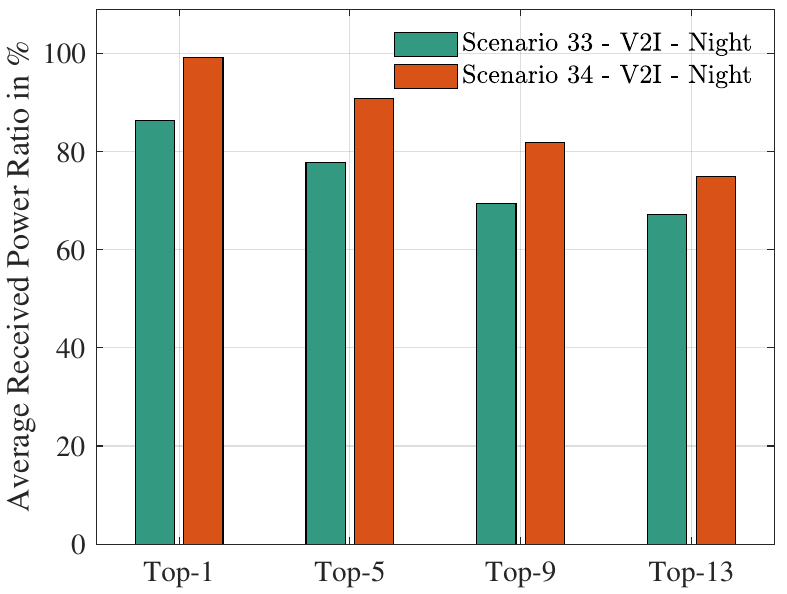}
    \includegraphics[width=0.24\textwidth]{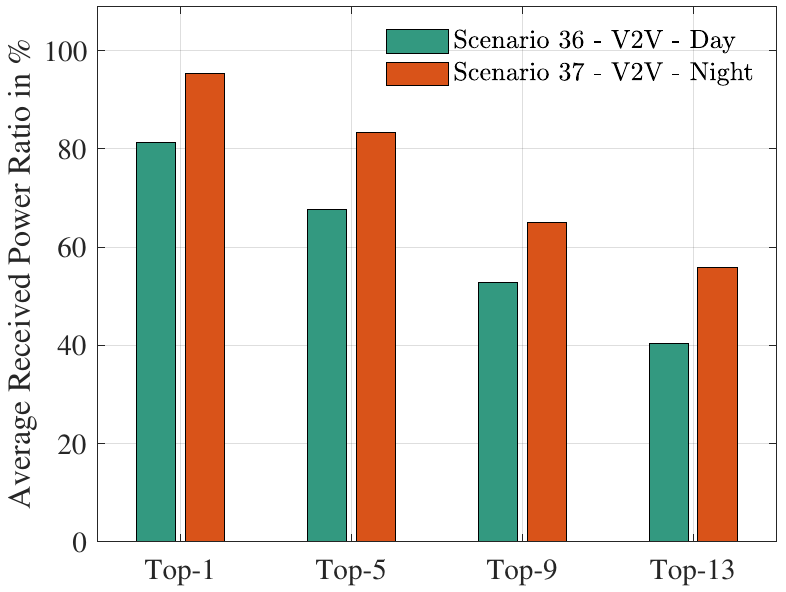}
    \includegraphics[width=0.24\textwidth]{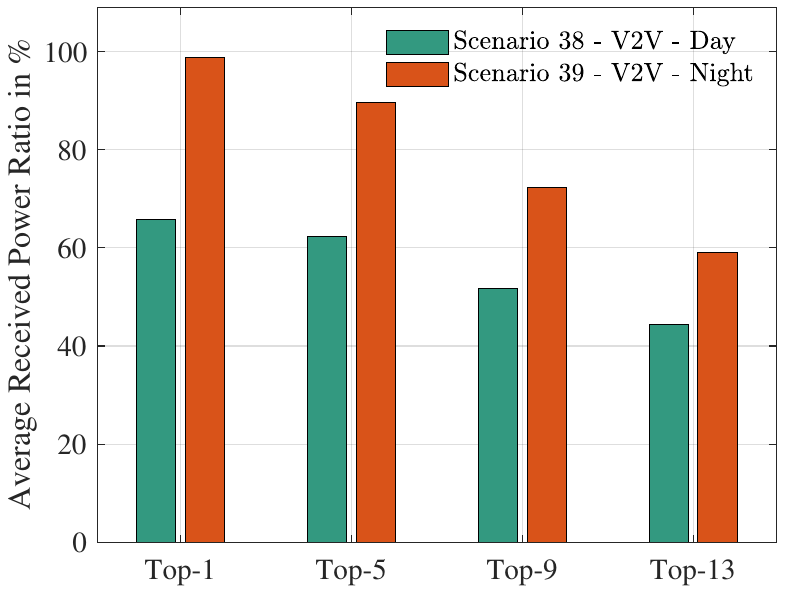}
	\caption{\centering \scriptsize Average achievable received power ratio on the testing data for all considered scenarios.}
\end{subfigure}
    \caption{The performances of average achieved top-$M$ accuracies and received power ratios in percentages tested on the vehicle-to-infrastructure and vehicle-to-vehicle scenarios.}
	\label{fig: Performances}
\end{figure*}

    For better analysis and illustration of the performance evaluation of the proposed solution, we consider following two matrices to measure the performances in terms of testing. Both matrices are averaged over total test samples. Besides, it should also be noted that all average results obtained from this experiments after running $5$ times.

    • \textit{Top-$M$ Accuracy:} The results of number of correct top-$M$ predictions from the model during test in relation to the total number of predictions can be defined as follows.
    
\begin{equation}
    Acc_{top-M} = \frac{1}{N_{test}}\sum_{t=0}^{N_{test} - 1}\frac{\mid I^*_t \cap \hat{I}_t \mid}{\mid \hat{I}_t \mid},
\end{equation}
    where, $N_{test}$, $I^*_t$, and $\hat{I}_t$ denote the number of test samples, optimal beams at $t^{th}$ time, and top-$M$ beams at $t^{th}$ time, respectively. 

    • \textit{Received Power Ratio:} The ratio between received power in downlink associated with predicted beams and received power from ground-truth beams is a measure of received power that can be possibly received from the link. It can be defined as follows.

\begin{equation}
    R_{\mathcal{P}_t} = \frac{1}{N_{test}}\sum_{t=0}^{N_{test} - 1}\frac{\hat{\mathcal{P}_t}}{\mathcal{P}_t^{gt}},
\end{equation}
    where, $\hat{\mathcal{P}_t}$ and $\mathcal{P}_t^{gt}$ are the received power from predicted beams and ground-truth beams, respectively, at time instant $t$.
    
    The results in terms of top-$M$ beam selection accuracies for eight scenarios are depicted in Fig. \ref{fig: Performances}(a) where the top beam candidates are $M \in \{1, 5, 9, 13\}$. It can be noticed that the accuracies keep on increasing as $M$ until the $M = 13$. For instance, while measuring the top-$13$ accuracies, it achieves up to $98.19\%$ and $66.96\%$ on V2I and V2V scenarios, respectively. For another, we also depict the results of received power ratio performances in Fig. \ref{fig: Performances}(b), which is our emphasis of this proposed solution. From the results, we can particularly observe that our proposed solution performs very well while also accomplishing sufficient maximum received power from the mmWave communications links as desired. For example, according to the results, with top-$1$ predicted beams, $99.92\%$ and $98.86\%$ received power ratio can be possibly achieved on average in V2I and V2V scenarios, respectively.  
    
    However, even though the aforementioned performance results are obtained from measured data in urban settings during day and night times, addressing the adaptability of our work to suburban or rural scenarios is also necessary for real word applicability. Besides, the various weather conditions might impact the performance of sensing information, and hence play an important role. As such, we suggest that the model should be trained and updated continuously to get better generalization ability along with or without minor performance losses while considering the deployment in real world. Next subsection considers how the proposed solution might be integrated with 5G NR (New Radio) standard, and after integration, how it can contribute to reducing the beam sweeping latency and searching space overheads.

\subsection{Comparison with Exhaustive Search in 5G-NR}
    To meet the 5G-NR standard specifications \cite{salehi2024omni,tuninato2023comprehensive}, the beam selection procedure needs to incorporate a brute-force approach, where the transmitter and receiver nodes exhaustively explore the best beams with the strongest signal strength via sweeping of all beams predefined at the codebooks of directions. During the initial access, the nodes exchange messages to establish a suitable directional link. In particular, the sweeping process starts with sending a set of synchronizing signal (SS) blocks originating from the transmitter node in a sequential manner. At the same time, the receiver node tunes its antenna arrays along the predefined beam directions until the best beams are located from the signal strengths. The set of SS blocks (each specific beam assigned under each SS block) are grouped into a single SS burst with a length of $5$ ms, and repeated on every $20$ ms as presented in Fig. \ref{fig: Timing_5G_Proposed}(a). Considering up to $32$ number of SS blocks fit into a single burst, we can express the total time requirements for beam selection process as follows:

\begin{figure*} [!t]
	\includegraphics[width=.95\linewidth]{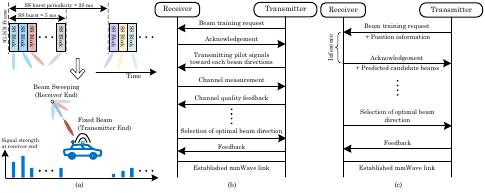}
    \centering
    \caption{(a) The frame structure of synchronizing signal (SS) burst within 5G-NR standard (each SS block corresponding to specific beam is represented as different colors) and (b)\&(c) the timing diagrams of 5G-NR standardized exhaustive beamforming and our approach with proposed multi-modal learning.}
    \label{fig: Timing_5G_Proposed}
\end{figure*}
    
\begin{equation}
    \mathcal{T}_{nr}(|\mathbfcal{Q}|) = T_{ssb} \left\lfloor\dfrac{|\mathbfcal{Q}|-1}{32}\right\rfloor + t_{burst},
\end{equation}

    where, $T_{ssb}$, $t_{burst}$, and $|\mathbfcal{Q}|$ are the duration of a periodicity, single burst, and number of beams, respectively. We can also get a SSB duration, i.e., a single beam sweep duration as $t_{ssb}$ = $t_{burst}/32$ = $0.156$ ms. However, the proposed solution seeks to reduce the number of beams to be explored from $|\mathbfcal{Q}|$ to identify predicted top-$M$ beam candidates. Then, the time associated with sweeping the predicted beams can be given by:

\begin{equation}
    \mathcal{T}_{nr}^{cnn}(M) = T_{ssb} \left\lfloor\dfrac{M-1}{32}\right\rfloor + t_{ssb}(1+(M-1)\;\mathrm{mod}\;32).
\end{equation}
    
    Fig. \ref{fig: Timing_5G_Proposed}(b) illustrates how 5G-NR defined exhaustive search beamforming works, whereas, Fig. \ref{fig: Timing_5G_Proposed}(c) represents how our proposed approach integrates with such standard defined existing beam searching approach to find the best beam directions for communications between two units. In particular, the proposed approach enables the receiver base station/road-side unit (V2I) or vehicle (V2V) to predict the candidate beams for the connected vehicles who wish to communicate with them. After that, 5G-NR standard based beam selection procedures can be employed, where both receiver and transmitter nodes systematically perform beam sweeping on the candidate beams over a range of pre-defined angular directions to find the most desired optimal directions. It should be noted that the beams are required to adjust according to the channel conditions and vehicular mobility over time. Hence, the procedure essentially involves measuring and giving feedback about the channel quality for each candidate beams. Upon analyzing the channel quality and feedback, the best beam direction can be determined eventually which in turn establish the mmWave communication link. 
    
\begin{figure} [!ht]
	\includegraphics[width=.85\linewidth]{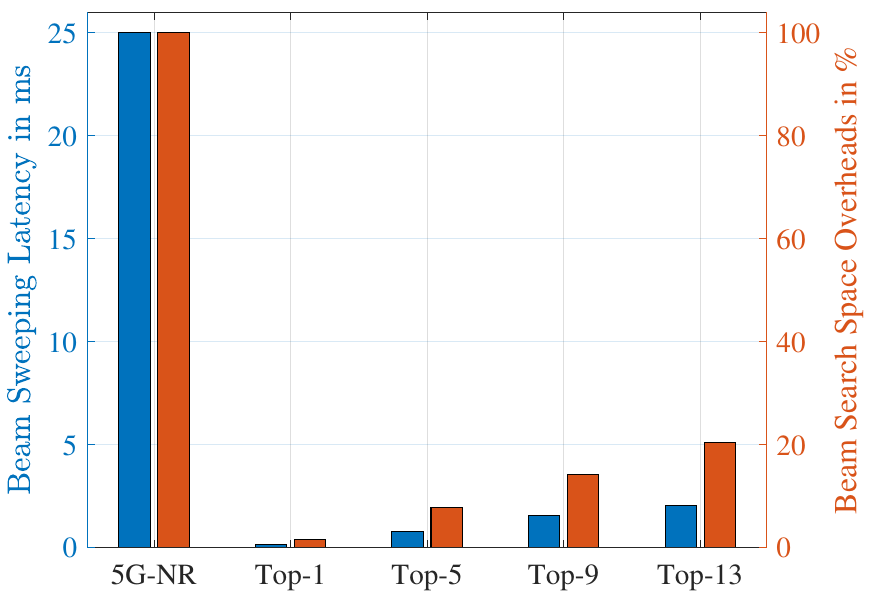}
    \centering
    \caption{Performance comparisons of standard defined existing beam selection approach and our proposed approach with respect to beam sweeping latency and beam searching space overheads.}
    \label{fig: Latency_Overheads}
\end{figure}

    In Fig. \ref{fig: Latency_Overheads}, we compare the performances of our proposed approach against standard defined existing beam selection approach in terms of beam sweeping latency and beam searching space overheads. The beam sweeping times are calculated from the Eqs. (14) and (15). As shown, the proposed approach requires up to $2.03$ ms to search top-$13$ candidate beams, a significant improvement over $25$ ms needed otherwise for searching all possible $64$ beams with existing exhaustive search. On the other hand, the beam search space overheads directly depend on the number of beams to be swept. For example, if we need to explore from top-$13$ predicted beams, the maximum searching overheads will be $13$ number of beams. Since the proposed approach considered in this work provides top-$M$ predicted beams to be explored, it could considerably shorten the beam searching overheads roughly to $20.31$\% when compared to exhaustive beam searching.
    
    In a sum, we have employed deep learning on out-of-band sensor information for beam selection to avoid unnecessary beam searching overheads by providing a selected number of predicted candidate beams. Such reductions of beam searching space is particularly desirable for vehicular settings since it eventually enables to shorten the communications delays, and possibly integrate also with 5G-NR standard. For instance, our evaluation results justify that the proposed solution greatly shortens beam sweeping searching and time overheads roughly by $79.67$\% and $91.89$\%, respectively for top-$13$ beams. However, despite the top-$1$ having minimal overheads, its low accuracy may mislead to identify the accurate best beam directions for communications. We can recall that we can get maximum accuracy from the model is $98.19$\% for top-$13$ predicted outcomes. As such, the simple intuition is employing top-$13$ beams would perform perfectly in predicting the correct beams, thereby ensuring to avoid possible beam misalignment with still very modest overheads. On the other hand, it is worth mentioning that while it takes longer time for training the proposed model, the average inference execution time (per sample) measured on NVIDIA GeForce RTX 2080 GPU is roughly between $0.5$ ms and $1.0$ ms, whereas the multi-modal data pre-processing times are negligible. This work is thus able to meet real-time requirements and be deployable in real world scenarios.
    
    
\section{Conclusion}
    In this paper, we have introduced a 5G-NR compatible beam selection solution to improve on harnessing the downlink received power in $60$ GHz mmWave enabled V2I and V2V communications. For that, we have presented a proposed multi-modal feature extractions architecture utilizing deep learning models to learn from the out-of-band multi-modality sensing information and predict the top-$M$ beams (a subset of beams). Subsequently, these predicted beams in turn helps to reduce the beam searching space, thereby addressing the beam searching overheads limitation of mmWave communications. In the end, experiments on real-world measured wireless datasets have shown the proposed solution's effectiveness and applicability, indicating the benefits of exploring multi-modality sensing in mmWave beamforming for connected vehicles. Perhaps, exploring how to incorporate multi-modality sensing and communications in non-line-of-sight (NLoS) scenarios within mmWave vehicular communications could be considered as a future work.

\section*{Acknowledgments}
    The authors would like to thank to the editor and reviewers for their thoughtful feedbacks toward improving the quality of this work.

\ifCLASSOPTIONcaptionsoff
  \newpage
\fi

\bibliographystyle{IEEEtran}
\bibliography{bibliography.bib}

\begin{IEEEbiography}[{\includegraphics[width=1in,height=1.25in,clip,keepaspectratio]{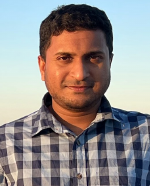}}]
    {Muhammad Baqer Mollah} is a PhD candidate in the Department of Electrical and Computer Engineering at the University of Massachusetts Dartmouth. Before joining at University of Massachusetts Dartmouth, he has been working at Nanyang Technological University (NTU) as a full-time Research Associate. His research interests include advanced communications, security, and edge intelligence techniques for Internet of Things (IoT) and connected vehicles. He has a M.Sc. in Computer Science and B.Sc. in Electrical and Electronic Engineering from Jahangirnagar University, Dhaka and International Islamic University Chittagong, Bangladesh, respectively.
\end{IEEEbiography}

\begin{IEEEbiography}[{\includegraphics[width=1in,height=1.25in,clip,keepaspectratio]{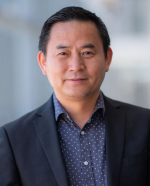}}]
    {Honggang Wang} is the founding department chair and Professor of the Department of Graduate Computer Science and Engineering at Katz School of Science and Health, Yeshiva University. He was a Professor in the Department of Electrical and Computer Engineering, University of Massachusetts Dartmouth. He was the “Scholar of The Year” awardee in 2016, the highest research recognition at University of Massachusetts Dartmouth. He has served as the Editor-in-Chief of IEEE Internet of Things Journal (2020 - 2022). He has won several prestigious best paper awards throughout his career. He was the past Chair of the IEEE Multimedia Communications Technical Committee (2018-2020) and the IEEE eHealth Technical Committee Chair (2020-2021). His recently focused research topics involve AI and Internet of Things, mmWave communications, connected vehicles, smart health, and cyber security. He has a PhD in Computer Engineering from University of Nebraska - Lincoln.
\end{IEEEbiography}

\begin{IEEEbiography}[{\includegraphics[width=1in,height=1.25in,clip,keepaspectratio]{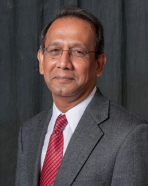}}]
    {Mohammad Ataul Karim} is a Professor in the Department of Electrical and Computer Engineering at the University of Massachusetts Dartmouth. His areas of research encompass optical computing, pattern/target recognition, computer vision, and Internet of Things. He is the former Provost, Executive Vice Chancellor for Academic Affairs, and Chief Operating Officer at University of Massachusetts Dartmouth. He is author/editor of 19 books, over 365 research papers, 15 book chapters, and 3 US patents. He serves on the editorial boards of Optics and Laser Technology, International J. of Modern Physics B, and Modern Physics Letters B. He received his PhD in electrical engineering degree from the University of Alabama. Karim is a fellow also of Optica, Society of Photo-Instrumentation Engineers, Institute of Physics, Institution of Engineering and Technology, Bangladesh Academy of Sciences, and Asia-Pacific Artificial Intelligence Association.
\end{IEEEbiography}

\begin{IEEEbiography}[{\includegraphics[width=1in,height=1.25in,clip,keepaspectratio]{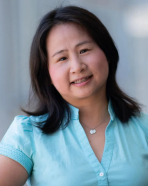}}]
    {Hua Fang} is a Professor in the Department of Computer and Information Science at the University of Massachusetts Dartmouth. She has been leading Computational Statistics and Data Science Lab which aims at developing computational methods, tools and systems for broader health and behavior studies by integrating and advancing the theories and applications of computational statistics, which is a computational science at the interface of statistics and computer science. Her current research interests include real-time machine or statistical learning, and visual analytics of wearable biosensor data streams, broadly in E-/M-/Digital/Tele-/Virtual/Precision health, and Internet of Things. 
\end{IEEEbiography}

\end{document}